\documentclass[a4paper,12pt]{article}
\usepackage[utf8]{inputenc}
\usepackage{cancel}
\usepackage{ulem}
\usepackage{amsfonts}
\usepackage{amssymb}
\usepackage{graphicx}
\usepackage{amsmath}
\usepackage{enumerate}
\usepackage{mathtools}
\usepackage{subfig}
\usepackage{color}
\usepackage{float}
\usepackage{here}
\usepackage{cite}
\usepackage{mathrsfs}
\usepackage{float,epsfig}
\usepackage{dcolumn}

\usepackage{graphicx}
\usepackage{bm}
\usepackage{amsmath,amssymb,amsthm}
\usepackage[colorlinks=true,linkcolor=blue,citecolor=red]{hyperref}
\usepackage[toc,page]{appendix}
\newcommand{\be}{\begin{equation}}
\newcommand{\ee}{\end{equation}}
\newcommand{\bea}{\setlength\arraycolsep{2pt} \begin{eqnarray}}
\newcommand{\eea}{\end{eqnarray}}

\def\0{{\sst{(0)}}}
\def\1{{\sst{(1)}}}
\def\2{{\sst{(2)}}}
\def\3{{\sst{(3)}}}
\def\4{{\sst{(4)}}}
\def\5{{\sst{(5)}}}
\def\6{{\sst{(6)}}}
\def\7{{\sst{(7)}}}
\def\8{{\sst{(8)}}}
\def\sst#1{{\scriptscriptstyle #1}}

\makeatletter \@addtoreset{equation}{section}

\usepackage{multirow}
\begin{document}
	
\title{\bf \Large
	Deflection  Angle and Shadow  Behaviors  of  Quintessential  Black Holes in arbitrary Dimensions }
\author{   A. Belhaj$^{1}$\footnote{belhajadil@fsr.ac.ma},  M. Benali$^{1}$, A. El Balali$^{1}$,  H. El Moumni$^{2}$\thanks{hasan.elmoumni@edu.uca.ma}, S-E. Ennadifi$^{3}$, \footnote{Authors in alphabetical order.}
	\hspace*{-8pt} \\
	{\small $^1$ D\'{e}partement de Physique, Equipe des Sciences de la mati\`ere et du Rayonnement,
		ESMaR}\\ {\small   Facult\'e des Sciences, Universit\'e Mohammed V de Rabat,  Rabat, Morocco} \\
	{\small $^{2}$  EPTHE, Physics Department, Faculty of Science,  Ibn Zohr University, Agadir, Morocco} \\  {\small  $^3$  D\'{e}partement de Physique,   Facult\'{e}
		des Sciences, Universit\'{e} Mohammed V de   Rabat,
		Morocco }
} \maketitle

	\begin{abstract}
		{\noindent}
Motivated by M-theory/superstring inspired models, we  investigate certain behaviors of the  deflection angle and the   shadow  geometrical shapes of higher dimensional  quintessential  black holes associated with  two values  of   the  dark energy (DE) state parameter, being $\omega=-\frac{1}{3} $ and  $\omega=-\frac{2}{3} $.  Concretely,  we derive the geodesic equation of photons on such  backgrounds. Thanks to  the Gauss-Bonnet theorem  corresponding to    the optical metric,  we   compute  the leading terms of the deflection angle in the so-called  weak-limit approximation.
After that, we inspect the  effect of DE
and  the  space-time dimension $d$
on  the calculated  optical  quantities. Introducing DE via  the  field intensity $c$ and  the state parameter $\omega$, we find  that  the shadow size and the deflection angle  increase  by  increasing values of the field intensity  $c$.   However, we observe that  the high dimensions decrease such  quantities for $\omega$-models exhibiting   similar  behaviors. Then, we consider  the effect of  the  black hole charge,  on  these  optical  quantities,   by discussing  the associated behaviors. The present investigation   recovers certain known results appearing in   ordinary  four dimensional models.
	\end{abstract}
\newpage	
	\tableofcontents
\newpage
	\section{Introduction}
Recently, after the successful observation of the first image of
the black hole in the electromagnetic spectrum in the center of galaxy M87
\cite{1,2}, there is a continuous improvement of measurements for much
higher resolution in the future \cite{3}, since a clear geometrical
identification of the black hole from the first image is not allowed. Subsequently, it is
hard to put aside the theoretical efforts dealing with  the black hole physics in
diverse gravitational theories and astrophysical mediums.

 Black holes involve the natural
peculiarity of sucking in surrounding matter in a phenomenon called
"accretion". This   matter,  accrediting on the black hole,  passes  through
the horizon. It has been observed   that this  gives a dark area on a light background
called
the "shadow" of the black hole being
based on the so-called gravitational lensing phenomenon. It turns out  that  the  shape and  the size of the light
generated by
matter,  flowing at the edge of the event horizon,  can be   determined by  the black hole parameters including
the mass
and rotation \cite{de2000apparent}.  For non-rotating black holes,  it has been  found that the  shape of the
shadow develops  a  circular geometry. However,  rotating black holes involve non trivial  shapes depending  on  the  rotation parameter  \cite{chandrasekhar1985mathematical}.
 Various black hole models, in arbitrary dimensions, have
been investigated using different methods and approaches including
supergravity theories \cite{singhshadow}. These activities have been
motivated by string theory requiring more than four physical directions.
In
fact, the first successful numerical counting of the entropy of
black-hole
in such a non-trivial theory was performed in five-dimensional
space-time
unveiling the microscopic stringy description of black holes
\cite{SV,ER}.
Moreover, several  efforts have been deployed to test extra dimensions in
future colliders, where the study of higher-dimensional black holes
properties could take place. In particular, the investigation of the
presence of extra dimensions greatly  could suggest   a new physics dealing with higher dimensional black holes. It is expected that this could be
experimentally accessible and manifests itself through a number of
strong
gravity effects, as soon as the energy of a given experiment exceeds
some
characterising fundamental scale under the the four-dimensional Planck
scale.
The Large Hadron Collider (LHC) at CERN, with a center-of-mass energy of
14 TeV,
becomes then a natural place to look for such  extra dimensions as well
as strong
gravity effects which could be associated with such  black holes.
Indeed, the latters, might be detected via their radiation spectra
according
to the evaporation process along with an additional number of distinct
observable signatures, which could support the  existence of these
extra
dimensions. The presence of such  hidden  dimensions has  became a  primordial  element in the comprehension of unified theories including  string/M-theory inspired models.  This could suggest that  there can be some remnants of the extra dimensions in the detection of the gravitational waves,  encoding some information  associated with  the underling  size  and the dynamics of fluctuations modes. In this  way,  many   works have been elaborated to unveil such  a physics  \cite{Cardoso:2016rao,Yu:2016tar,Visinelli:2017bny,Kwon:2019gsa}. Despite many attempts to support such activities, it has not been successful up to now. Nowadays,  thanks to EHT, we dispose of  new tools  to continue hunting of the   hidden dimensions \cite{Vagnozzi:2019apd}.
 However,  in order   to understand  the observed black hole shadow, gravitational lensing can be a helpful instrument of astrophysics  and astronomy \cite{Perlick:2004tq, Perlick:2018iye}.  Precisely,  the discovery of dark matter filaments with the help of  the weak deflection is an extremely relevant topic since it is very helpful in the investigation of the   Universe structure \cite{Epps:2017ddu,Bartelmann:2016dvf,Bambi:2019tjh,Allahyari:2019jqz,Cunha:2016wzk,Shaikh:2018lcc,Tao,hajafatima}.

Besides, it is now widely supposed that only 5\% of
mass and energy in the universe is visible and  it is well described within the
standard model of particle physics \cite{4}. While,  the remaining large
hidden
part consists of 25\% of dark matter and 70\% of dark energy (DE), whose the
existence and  the nature are still undetermined  \cite{5,6}. The supposed
existence
of dark matter is highly motivated by the non-Newtonian behavior of high
velocities of stars at the outskirts of galaxies. This might  imply that
visible disks of galaxies are flooded in a much larger, roughly
spherical, halo of invisible matter \cite{7,8}. For DE, affecting the
universe on the largest scales, the first observational proof for its existence
arose
from supernovae measurements. Namely, distant Ia-type supernova
explosions point out that a very small repulsive cosmological constant, i.e., vacuum energy, quintessential field,
manifesting repulsive gravitational effect,   are needful for the
explanation
of the accelerated expansion of the recent Universe \cite{9,10}. Similar
motivations in favor of DE are also concluded by the Planck
space
observatory measurements of the cosmic microwave background \cite{11}.

Since the role of the vacuum energy  has been  widely discussed in cosmological
models \cite{12,14,51}, thus, it is also pertinent to consider its role
in the physical processes taking place near black holes, essentially in the
vicinity of
the black hole horizon \cite{VV}.  More recently,   the physics of  black holes in the presence of such an energy  has  been extensively developed. Precisely, a special emphasis has put on   quintessential black holes  from  M-theory/superstring inspired models \cite{boby1}.

The aim of this  work is  to  contribute to these activities by  investigating
certain optical   behaviors  of   higher dimensional  quintessential
black hole (QBH) and estimating  the energy emission rate  associated with  two values of   the   DE  state parameter being $\omega=-\frac{1}{3} $ and  $\omega=-\frac{2}{3} $.    Concretely,  we  get  the geodesic equations of photons on such  backgrounds.  Using   the Gauss-Bonnet theorem associated with   the optical metric,  we   calculate   the leading terms of the deflection angle in the weak-limit approximation  framework.
Varying the  space-time dimension $d$ and introducing DE via  the  field intensity $c$ and  the state parameter $\omega$, we find  that  the shadow size and the deflection angle increase  within increasing values of the field intensity  $c$.   However,  it has been shown  that  the higher dimensions decrease such  quantities for $\omega$-models  revealing    similar  behaviors. Then, we  study   the effect of  the  black hole charge  on  such computed  optical  quantities. The  present investigation,   which   recovers   some  known four dimensional  ordinary  results, comes up  with certain open questions associated with M-theory/superstring inspired models where DE could find a possible place supported by  extra dimensions.

The paper is organized as follows. In  section 2,  we reconsider the study of
the  non-charged Schwarzschild-Tangherlini solutions in higher dimensions with the
presence of quintessential DE. In particular, we investigate the effective potential and  the  shadow  geometrical behaviors of such a QBH.  In section 3,  we  compute and graphically analyse, in
some
details, the significant impact of  the quintessential energy on the
Shwarzschild-Tanglerlini black hole deflection angle. In this section 4, we study
the
effect of  the charge  of  Schwarzschild-Tangherlini holes on such optical  quantities. The section 5
is
devoted to the discussion on possible extensions,  the  summary of the work,  and certain open questions.  Some used material associated with the metric calculations are given in the appendix.
\section{Shadow behaviors of Quintessential Schwarzschild-Tangherlini black hole}
In this section, we reconsider the investigation of  quintessential Schwarzschild-Tangherlini black holes \cite{tangherlini_schwarzschild_1963}.  Before going ahead,  it is now known that the general properties of the universe
are
described by assuming that its dynamics are ruled by an energy source,
i.e., DE, whose the physical origin remains unknown. This
cosmological
component has an energy-momentum tensor which  can be obtained from
Friedmann's equations.  A remarkable characteristic of this
antigravitational energy component is its negative pressure which is
comparable, in absolute value, to the energy density. Therefore,
whatever
its nature is,  DE can be effectively depicted in terms of
the
pressure and  the density.  Treated as a perfect fluid with pressure $p$
and energy density $\mathbf{\rho }$,  a
parametrization
of DE is possible via the introduction of what is known as the
equation of state parameter, being the ratio of its pressure
and
density
\begin{equation}
\label{R3}
\omega=\frac{p}{\mathbf{\rho}}.
\end{equation}%
From such an equation of  the state, some of the well studied cases
of fluids are
\begin{itemize}
 \item $\omega=-1$  associated with  the cosmological constant $\Lambda $,
\item $\omega=0$  corresponding to  a pressureless regime like non-relativistic
matter,
i.e., dust
\item  $\omega=1/3$ associated with  a radiation.
\end{itemize}
For a repulsive gravity effect, it appears that, in a
homogeneous
and isotropic universe, the corresponding fluid equation of state is $\omega<-1/3
$.  Therefore,  the cosmological constant $\Lambda $, or
any
fluid with equation of state $\omega<-1/3$ accelerates the
expansion.
One of such a hypothetical DE  fluid is the quintessence.   The
latter
is a dynamical, evolving, spatially inhomogeneous component  (unlike a
cosmological constant, its pressure and energy density evolve in time). Thus,  $\omega $ may also do so  with equation of state
$-1<\omega<0.$
The smaller is the value of $\omega$, the greater its accelerating
effect. Such a dynamical  DE  field which is thought to drive the
overall cosmic history of the universe, may also still affect its large
structure, for instance, galaxies, black holes including their thermodynamical and optical  aspects.\\
 In connections with gravity models, there  has been a significant interest in the study of  higher dimensional  of Einstein equations providing  black hole solutions   considered as the  most exact ones  of general relativity. Such solutions could encourage  the   black hole buildings from string theory involving more than four dimensions. Such higher dimensional solutions, which will be dealt with, could be supported by the physics of extra dimensions being a possible investigation subject within future colliders including  LHC. Motivated by such activities, we would like  to study the shadow of the black hole in arbitrary  dimensions.

 According to \cite{bo11}, the line element of the metric,  in higher dimensional space-times with static black holes,   takes the following form
\begin{equation}
\label{R1}
ds^2=-e^\nu dt^2 +e^\lambda dr^2 + r^2d\Omega^2_{d-2},
\end{equation}
where  $ d\Omega^2_{d-2}=d\theta^2_1+\sin^2\theta_1^2d\theta^2_2+\ldots+\prod^{d-3}_{i=1}\sin^2\theta_id\theta^2_{d-2}$   represents the metric  on  the $ (d-2)$-dimensional unit sphere. It is noted that  $\nu$ and  $\lambda$  are two functions of the radial coordinate $r$,  respectively. In the  static spherically symmetric state,  in presence of the quintessence,  the the energy-momentum tensor can be written as \cite{VV}
\begin{equation}
{T_t}^t = A(r), \qquad {T_t}^j = 0,\qquad
{T_i}^j = C(r)\,r_i\,r^j +B(r)\,{\delta_i}^j.
\end{equation}
The average angle of the isotropic state provides
\begin{equation}
{\langle {T_i}^j\rangle = D(r)\,{\delta_i}^j,} \qquad
{D(r) = -\frac{1}{d-1}\,C(r)\,r^2 +B(r).}
\end{equation}
For the quintessence,  one has the solution
\begin{equation}
{D(r) = - w_q\, A(r)}
\end{equation}
where  $A(r)$ is  a  density term. It is recalled that,  for a fixed state parameter $\omega$,  we can get  the expression of $D(r)$. $C(r)\equiv 0$  represents  the condition associated with  a free quintessence \cite{VV}.  Taking  the metric Eq.(\ref{R1}) and applying  the calculations given  in the appendix, we  obtain  Einstein's equation. Indeed, the involved terms are
\begin{eqnarray}
{2T_{t}^{\;t}}\;&=&{\frac{(d-2) e^{-\lambda } \left((d-3) \left(e^{\lambda }-1\right)+r \lambda '\right)}{2 r^2},}\\
{2T_{r}^{\;r}}\;&=&\;{\frac{(d-2) e^{-\lambda } \left((d-3) \left(e^{\lambda }-1\right)-r \nu '\right)}{2 r^2},}\\
{2T_{\theta_i}^{\;\theta_i}}\;&=&\;{-\frac{e^{-\lambda } \left(2 (d-3) r \left(\nu '-\lambda '\right)+2 (d-4) (d-3)-r^2 \lambda ' \nu '+2 r^2 \nu ''+r^2 \left(\nu '\right)^2\right)}{4 r^2}}\\
 &+&{\frac{(d-3) (d-4)}{2 r^2}}\hspace{1cm} {i=1,\ldots, d-2,}
\end{eqnarray}
where the prime represent  the derivative with  respect  to $r$. In a higher dimensional spherically-symmetric space-time,   the general expression of the energy-momentum tensor  in the presence of the  quintessence reads as
\begin{eqnarray}
\label{R11}
&&{{T_t}^t} = {\rho_q(r),} \\[2mm]
\label{R22}
&&{{T_i}^j}  = { \rho_q(r)\,\alpha
\left[-(1+(d-1)\,B)\frac{r_i\,r^j}{r_n r^n}+B\,{\delta_i}^j\right]}.
\end{eqnarray}
It is noted that there  is a proportionality between the spatial components and the temporal one  with the arbitrary parameter $B$ depending on the internal structure of the quintessence. Considering  $\langle r_i\,r^j\rangle = \frac{1}{d-1}\, {\delta_i}^j\,r_n r^n$, the isotropic average over the angle results is given by
\begin{equation}
\left\langle{{T_i}^j}\right\rangle = {- \rho_q(r)\,\frac{\alpha}{d-1}\,
{\delta_i}^j = - p_q(r)\,{\delta_i}^j,}
\end{equation}
Using Eq.(\ref{R3}), one  has the constraint
\begin{equation}
\label{R2 }
w=\frac{\alpha}{d-1}.
\end{equation}
Applying the principle of  the additivity and the  linearity,   used in \cite{VV,bo11},    we get
\begin{equation}
\label{Lambda}
\lambda=-\ln(f_\omega),
\end{equation}
By the help of  Eq.(\ref{Lambda}), one obtains  the  linear differential equations involving  $f_\omega$
\begin{eqnarray}
\label{R33}
{T_{t}^{\;t} }& = & {T_{r}^{\;r}} =-\frac{d-2}{2r^2}(rf'_\omega+(d-3)(f_\omega-1)),\\
\label{R34}
{T_{\theta_i}^{\;\theta_i}} & = &{- \frac{1} {4r^2}(r^2f''_\omega+2(d-3)rf'_\omega+(d-4)(d-3)(f_\omega-1)),} \quad {i=1,\ldots, d-2.}
\end{eqnarray}
Combining equations Eq.(\ref{R11}), Eq.(\ref{R22}) and Eq.(\ref{R33}),  one can find   the fixed parameter $B$ given by
\begin{equation}
\label{ }
{B=-\frac{(d-1)\omega+1}{(d-1)(d-2)\omega}}.
\end{equation}
The energy-momentum tensor, appearing  in Eq.(\ref{R22}),  takes the following form
\begin{eqnarray}
\label{R31}
{T_{t}^{\;t} }& = & {T_{r}^{\;r}} =\textcolor{red}{\rho,}\\
\label{R32}
{T_{\theta_i}^{\;\theta_i}} & = &{- \frac{1} {d-2}(1+\omega(d-1)),} \quad {i=1,\ldots, d-2,.}
\end{eqnarray}
From  Eq.(\ref{R33}), Eq.(\ref{R34}), Eq.(\ref{R31}) and Eq.(\ref{R32}), we get  a differential equation for $f$
\begin{equation}
\label{R55}
{r^2f''_\omega+(d(\omega+2)-(\omega+5))rf'_\omega+(d-3)(f_\omega-1)(d(\omega+1)-(\omega+3))=0,}
\end{equation}
providing a   solution  given by
\begin{equation}
\label{7}
{f_\omega(r)=1-\frac{\mu}{r^{d-3}}-\frac{c}{r^{\omega(d-1)+d-3}},}
\end{equation}
where $\mu$ and $c$ are normalization factors. The energy density $\rho$, given in  Eq.(\ref{R3}),  should be positive and take the following form
\begin{equation}
\label{ }
{\rho=\frac{c(d-1)(d-2)\omega}{4r^{(d-1)(\omega+1)}}}.
\end{equation}
The line element of the  metric,  in  such a higher dimensional  spherically symmetric black hole surrounded by  the quintessence, reduces to
\begin{equation}
\label{2}
{ds^2=-f_\omega(r)dt^2+\frac{1}{f_\omega(r)}dr^2+r^2d\Omega^2_{d-2}},
\end{equation}
and $\mu$ is related to the  black hole masse $M$ through the relation
\begin{equation}
\label{4}
\mu=\frac{16\pi M}{(d-2)\Omega_{d-2}},
\end{equation}
where  one has $
\Omega_{d-2}=\frac{2\pi^{\frac{d-1}{2}}}{\Gamma(\frac{d-1}{2})}$. Roughly, the Lagrange and  the Hamilton-Jacobi equation  can be exploited to get  the equations of motion generating    QBH shadow geometric shapes using the following Lagrangian
\begin{equation}
\label{8}
\mathcal{L}=\frac{1}{2}g_{\mu\nu}\dot{x}^\mu\dot{x}^\nu.
\end{equation}
The solution of the  canonically conjugate momentums provides
\begin{eqnarray}
\label{12.1}
\frac{dt}{d\tau} =  \frac{E}{f_\omega(r)},\quad
\frac{d\theta_{d-2}}{d\tau} =  \frac{L}{r^2\prod_{i=1}^{d-3}\sin^2\theta_i},
\end{eqnarray}
where  $\tau$ is the affine parameter along the geodesics. $E$ and $L$  are   the energy and the angular momentum of the test particle, respectively.
To get  the shadow of the black hole,  one needs first  to obtain the geodesic form of such a  particle. To reach that,  the Hamilton-Jacobi equation and Carter constant separable method  should be used, matching the rotating case\cite{carter1968global,carter_global_1968}. Indeed, the  Hamilton-Jacobi equation  is expressed as
\begin{equation}
\label{12}
\frac{\partial{S}}{\partial \tau}=-\frac{1}{2}g^{\mu\nu}\frac{\partial S}{\partial x^\mu}\frac{\partial S}{\partial x^\nu},
\end{equation}
where $S$ is  the action Jacobi.   The separable solution  allows one to express the   action  as follows
\begin{equation}
\label{14}
S=\frac{1}{2}m^2_0\tau-Et+L\theta_{D-2}+S_r(r)+\sum_{i=1}^{D-3}S_{\theta_i}(\theta_i),
\end{equation}
where $m_0$ is the mass of the test particle.  $S_r(r)$ and  $S_{\theta_i}(\theta_i)$  are function depending  of $r$ and $\theta$, respectively.
Considering   a  test photon particle,  the calculation   provides
\begin{equation}
\label{17}
\begin{split}
0&=\bigg\{f_\omega(r)^{-1}\bigg(\frac{\partial S }{\partial t}\bigg)^2-f_\omega(r)\bigg(\frac{\partial S }{\partial r}\bigg)^2-\frac{1}{r^2\prod_{i=1}^{D-3}\sin^2\theta_i}\bigg(\frac{\partial S_{\theta_{D_2}}}{\partial \theta_{D-2}}\bigg)^2\bigg\}\\
&-\bigg\{\sum_{i=1}^{D-3}\frac{1}{r^2\prod_{n=1}^{i-1}\sin^2\theta_n}\bigg(\frac{\partial S_{\theta_{i}} }{\partial \theta_{i}}\bigg)^2\bigg\},
\end{split}
\end{equation}
In this way, the  separability of the equation  gives
\begin{equation}
\label{18}
\begin{split}
0&=\bigg\{f_\omega(r)^{-1}\bigg(\frac{\partial S }{\partial t}\bigg)^2-f_\omega(r)\bigg(\frac{\partial S }{\partial r}\bigg)^2-\frac{1}{r^2}\Bigg{(}\frac{1}{\prod_{i=1}^{D-3}\sin^2\theta_i}\bigg(\frac{\partial S_{\theta_{D-2}}}{\partial \theta_{D-2}}\bigg)^2\\
&+\mathcal{K}-\bigg(\frac{\partial S_{\theta_{D-2}}}{\partial \theta_{D-2}}\bigg)^2\;\prod_{i=1}^{D-3}\cot^2\theta_i\Bigg{)}\bigg\}\\
&-\bigg\{\frac{1}{r^2}\Bigg{(}\sum_{i=1}^{D-3}\frac{1}{\prod_{n=1}^{i-1}\sin^2\theta_n}\bigg(\frac{\partial S_{\theta_{i}} }{\partial \theta_{i}}\bigg)^2-\mathcal{K}+\bigg(\frac{\partial S_{\theta_{D-2}}}{\partial \theta_{D-2}}\bigg)^2\;\prod_{i=1}^{D-3}\cot^2\theta_i\Bigg{)}\bigg\},
\end{split}
\end{equation}
where $ \mathcal{K}$ is the Carter constant. Replacing  $\Big(\frac{\partial S_{\theta_{D-2}}}{\partial \theta_{D-2}}\Big)$   and $\big(\frac{\partial S }{\partial t}\big)$ by theirs expressions and introducing  $L$ and
 $E$, we obtain
\begin{equation}
\label{19}
\begin{split}
0&=\bigg\{f_\omega(r)^{-1}E^2-f_\omega(r)\bigg(\frac{\partial S }{\partial r}\bigg)^2-\frac{1}{r^2}\Bigg{(}\frac{L^2}{\prod_{i=1}^{D-3}\sin^2\theta_i}+\mathcal{K}-\prod_{i=1}^{D-3}L^2\cot^2\theta_i\Bigg{)}\bigg\}\\
&-\bigg\{\frac{1}{r^2}\Bigg{(}\sum_{i=1}^{D-3}\frac{1}{\prod_{n=1}^{i-1}\sin^2\theta_n}\bigg(\frac{\partial S_{\theta_{i}} }{\partial \theta_{i}}\bigg)^2-\mathcal{K}+\prod_{i=1}^{D-3}L^2\cot^2\theta_i\Bigg{)}\bigg\},
\end{split}
\end{equation}
 After simplifications,  we  obtain
\begin{eqnarray}
\label{20.1}
r^2f_\omega(r)^2\bigg(\frac{\partial S }{\partial r}\bigg)^2& = & E^2r^2-f_\omega(r)(\mathcal{K}+L^2)\\
\label{20.2}
\sum_{i=1}^{D-3}\frac{1}{\prod_{n=1}^{i-1}\sin^2\theta_n}\bigg(\frac{\partial S_{\theta_{i}} }{\partial \theta_{i}}\bigg)^2& = & \mathcal{K}-\prod_{i=1}^{D-3}L^2\cot^2\theta_i.
\end{eqnarray}
 Exploiting  \eqref{12.1},  \eqref{20.1},  \eqref{20.2} and the  definition of  canonically conjugate momentum,  we obtain the complete equations of motion for  photon ($m_0=0$)  around the Schwarzschild-Tangherlini  quintessential black hole
\begin{eqnarray}
\label{21.1}
\frac{dt}{d\tau}& = & \frac{E}{f_\omega(r)}\\
\label{21.3}
r^2\frac{dr}{d\tau}& = &\pm\sqrt{\mathcal{R}}\\
\label{21.4}
r^2\sum_{i=1}^{D-3}{\prod_{n=1}^{i-1}\sin^2\theta_n}\frac{d\theta_{i}}{d\tau}& = & \pm\sqrt{\Phi_i}\\
\label{21.2}
\frac{d\theta_{D-2}}{d\tau}& = & \frac{L}{r^2\prod_{i=1}^{D-3}\sin^2\theta_i}
\end{eqnarray}
where  the involved   quantities  $\mathcal{R}$ and $\Phi_i$ are given, respectively,  by
\begin{eqnarray}
\mathcal{R}(r)=  E^2r^4-r^2f_\omega(r)(\mathcal{K}+L^2), \quad
\Phi_i(\theta_i) =  \mathcal{K}-\prod_{i=1}^{d-3}L^2\cot^2\theta_i.
\end{eqnarray}
Indeed, the geometric  shape of a black hole is totally defined by the limit of its shadow  being the visible shape of the unstable circular orbits of photons.  To reach that,   one can use the radial equation of motion  which reads as
\begin{equation}
\label{22}
\Big(\frac{dr}{d\tau}\Big)^2+V_{eff}(r)=0,
\end{equation}
where $V_{eff}(r)$ is the effective potential for a  radial particle motion   given by
\begin{equation}
\label{23}
V_{eff}=\frac{f_\omega(r)}{r^2}(\mathcal{K}+L^2)-E^2.
\end{equation}
The maximal  value of the effective potential   corresponds to   the  circular orbits and  the unstable photons required by
\begin{equation}
\label{24}
V_{eff}=\frac{d V_{eff}}{d r}\Big|_{r=r_0}=0, \qquad  \mathcal{R}(r)=\frac{d\mathcal{R}(r)}{d r}\Big|_{r=r_0}=0.
\end{equation}
Using  \eqref{23}  and  \eqref{24}, we get
\begin{equation}
\label{25}
V_{eff}|_{r=r_0}=\frac{d V_{eff}}{d r}\Big|_{r=r_0}= \left\{
    \begin{array}{ll}
        &\frac{f_\omega(r_0)}{r_0^2}(\mathcal{K}+L^2)-E^2=0, \\
        \\
        &\frac{r_0f'_\omega(r_0)-2f_\omega(r_0)}{r_0^3}(\mathcal{K}+L^2)=0.
    \end{array}
\right.
\end{equation}
\subsection{Effective potential behavior}
\begin{figure}[!ht]
		\begin{center}
		\centering
			\begin{tabbing}
			\centering
			\hspace{9.3cm}\=\kill
			\includegraphics[scale=.5]{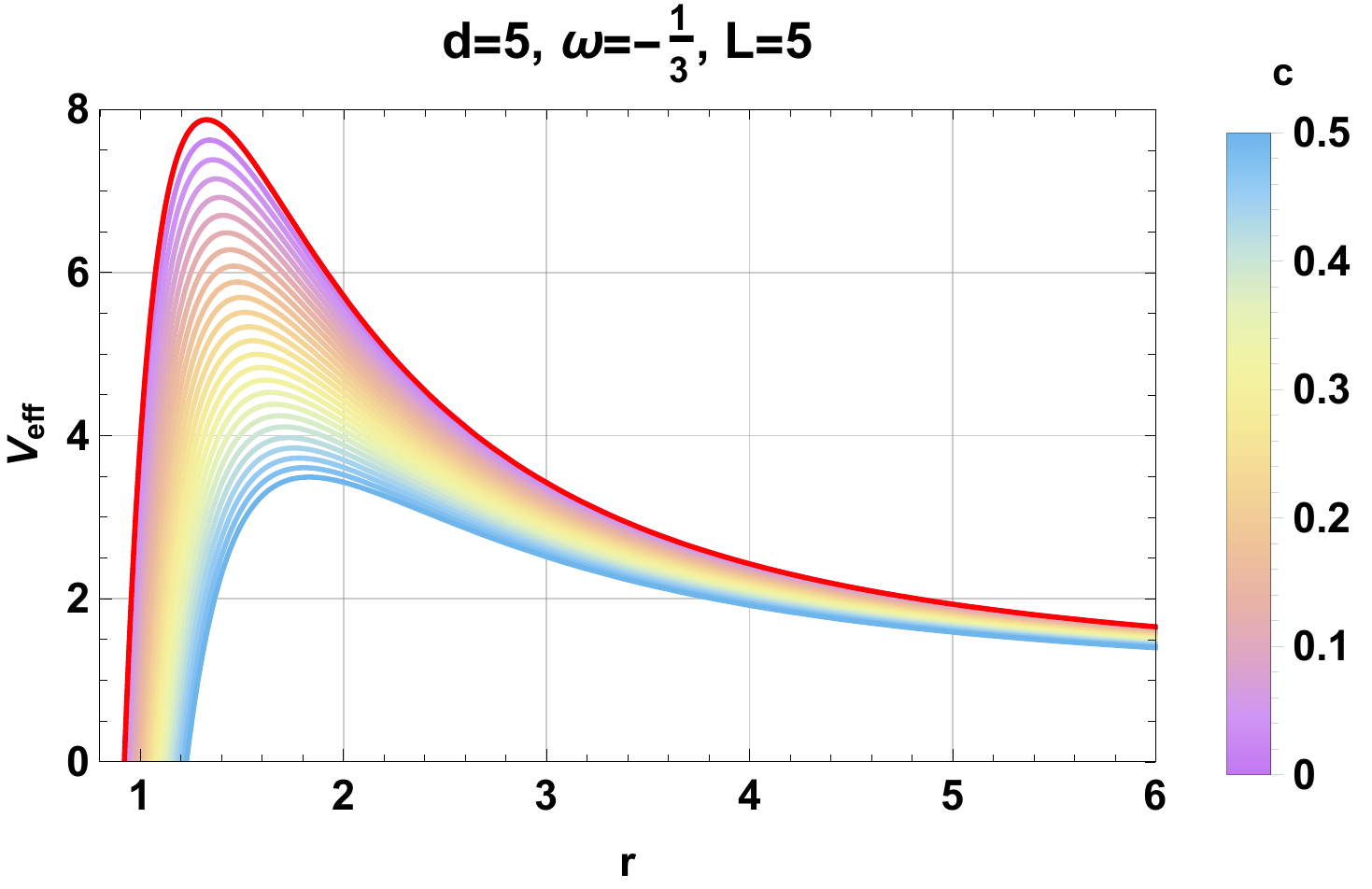} \>
			\includegraphics[scale=.5]{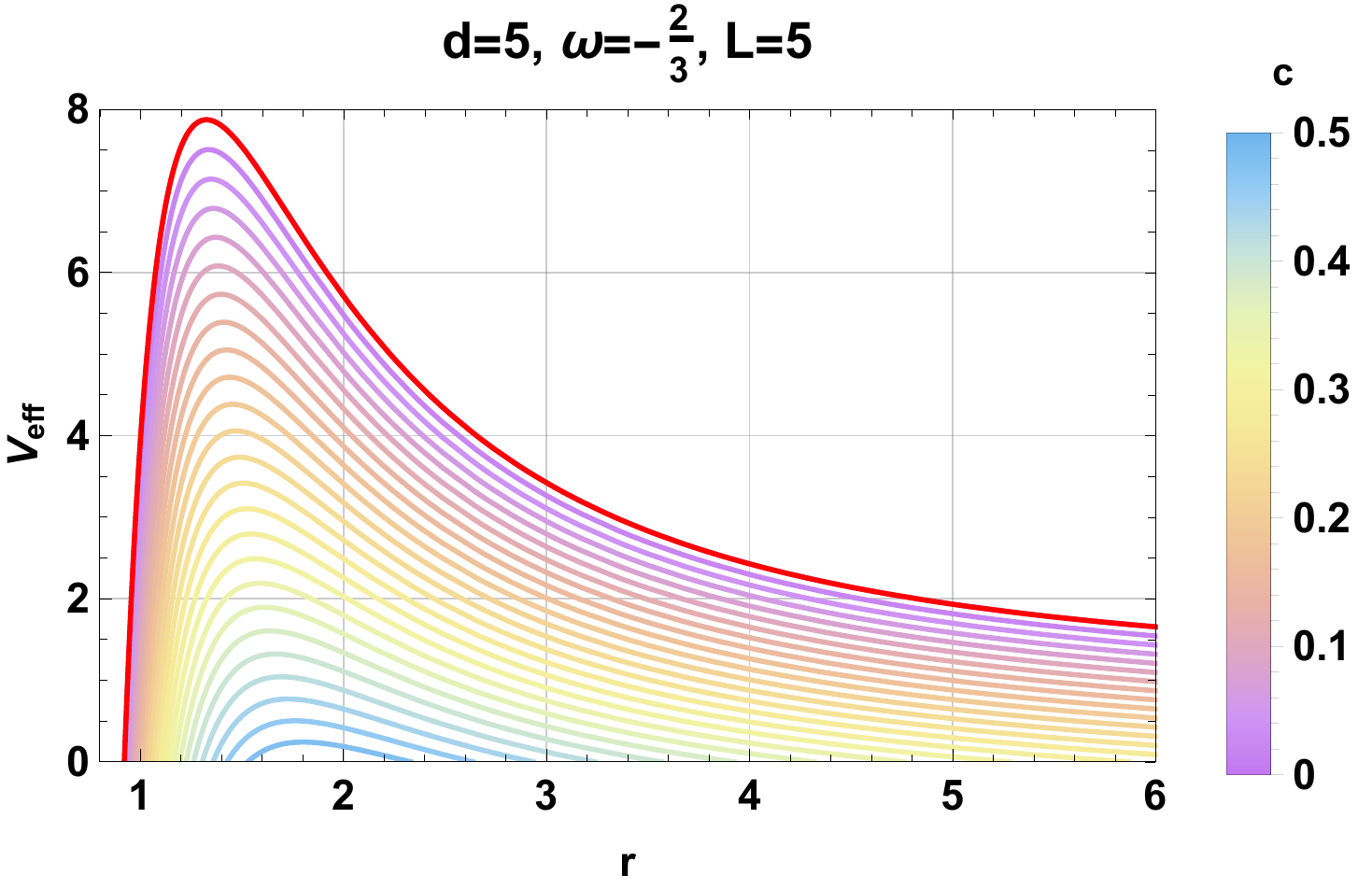} \\
			\includegraphics[scale=.5]{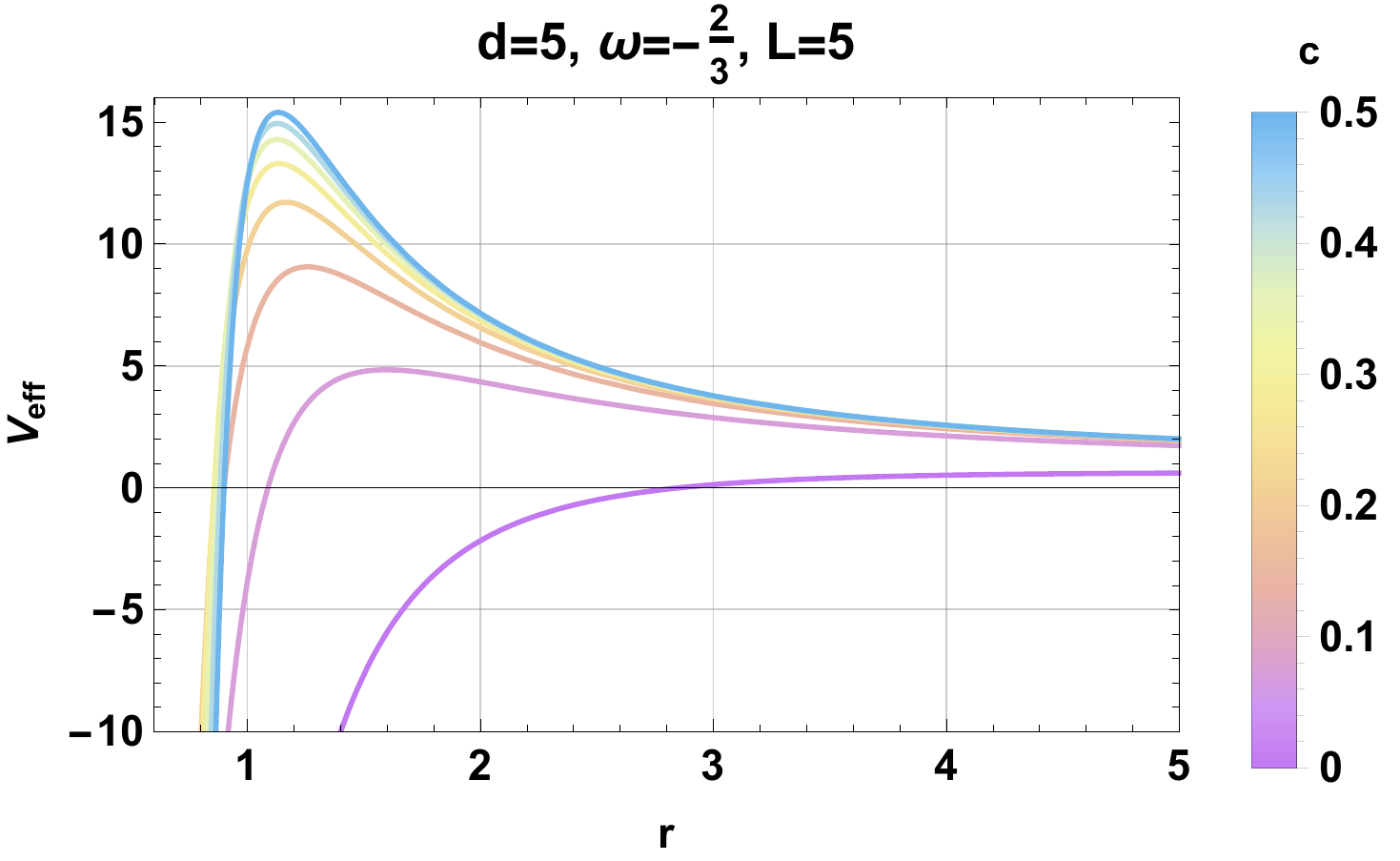} \>
			\includegraphics[scale=.5]{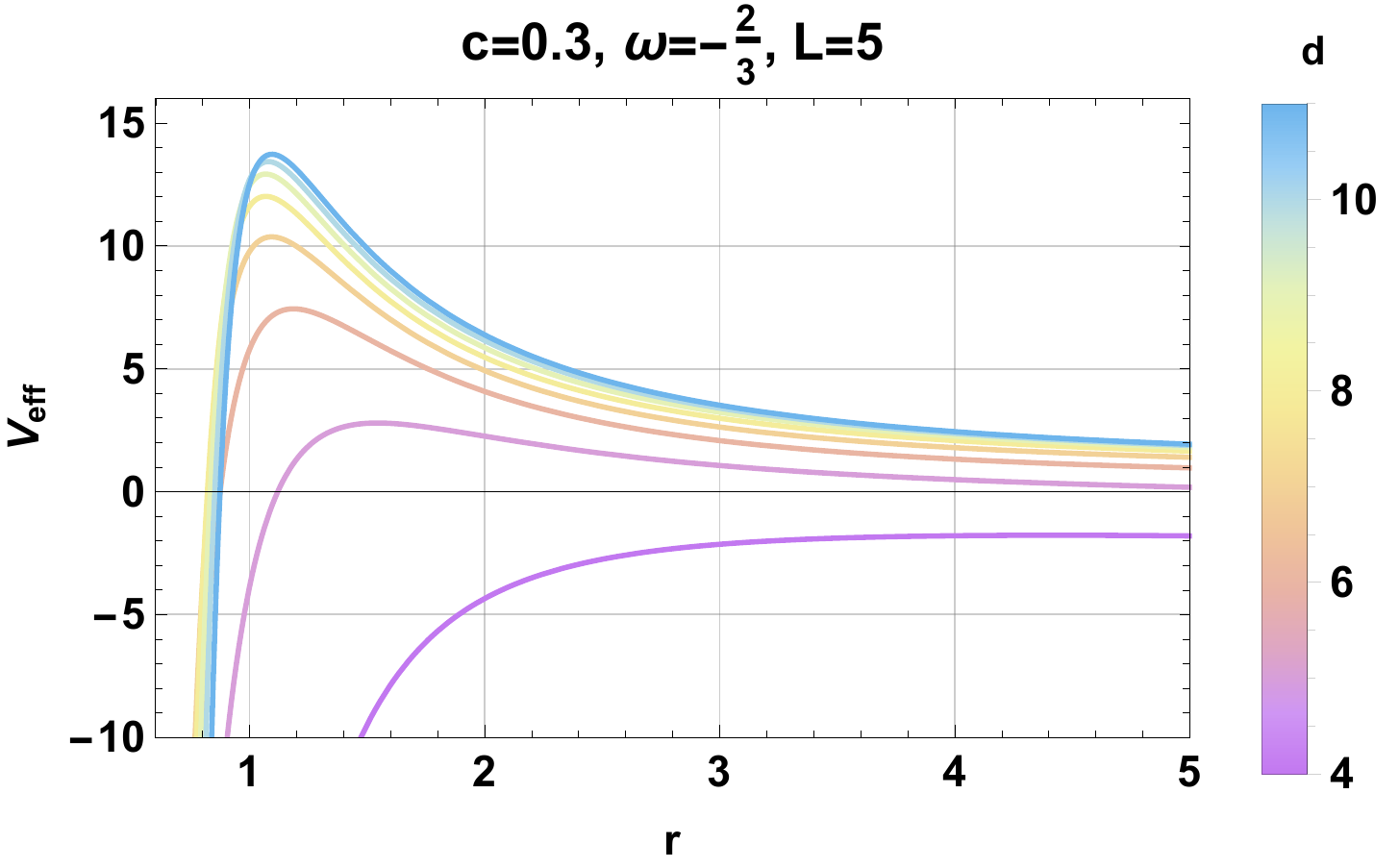} \\
		   \end{tabbing}
\caption{{\it \footnotesize The variation of the effective potential as a function of the radial coordinate in different dimensions $d$, $c$ and  two values of $\omega$, for  $L=5$,   $\mathcal{K}=1$ and $E=1$. In the all panels,  the red curve corresponds to the Schwarzschild solutions.}
}\label{f1}
\end{center}
\end{figure}

The effective potential of  the Schwarzschild-Tangherlini black holes with DE exhibits a   maximum for the photon sphere radius $r_0$  corresponding  to the real and  the positive solution of the following constraint
\begin{equation}
\label{26}
r_0f'_\omega(r_0)-2f_\omega(r_0)=0.
\end{equation}
To  analyse the  effective potential behaviors, we illustrate,  in  Fig.\ref{f1},  such a potential   as a function of the radial coordinate in  arbitrary  dimensions  $d$ and  $c$   for two different values of the  state parameter $\omega$, called in what follows the $(\omega)$-models.
 It  is worth nothing that  this matches perfectly  with the ordinary case  associated   with  the  Schwarzschild space-time  with a photon sphere radius $r_0=3M$ in the absence of DE.  It has been observed also  that  the  shadow boundary corresponds to a maximum effective potential value  being  almost the same  value   for   the DE  state parameter $\omega=-\frac{1}{3}$ and $\omega=-\frac{2}{3}$. However,   the  shadow boundary and  the  effective potential vary  in terms of  the radial coordinate  for  different dimensions $d$ and  $c$.
 It has been observed from  Fig.\ref{f1} that  the  potential for the Schwarzschild-Tangherlini black holes
is relevant than the one  surrounded  by  DE.
  Moreover,   the potential increases with the  dimension $d$. In this way,  the  unstable circular orbits become smaller. However,   the  $(-\frac{1}{3})$ and $(-\frac{2}{3})$-models exhibit  the same  maximum values showing  a universal behavior  with respect to such  an effective potential. Another important remark is that  the effective potential asymptote   is constant within the  large values of the radial coordinate $r$.
\subsection{Shadow behavior}
\begin{figure}[ht!]
		\begin{center}
		\vspace{-0cm}
		\centering
			\begin{tabbing}
			\centering
			\hspace{8.2cm}\=\kill
			\includegraphics[width=6.8cm, height=5.2cm]{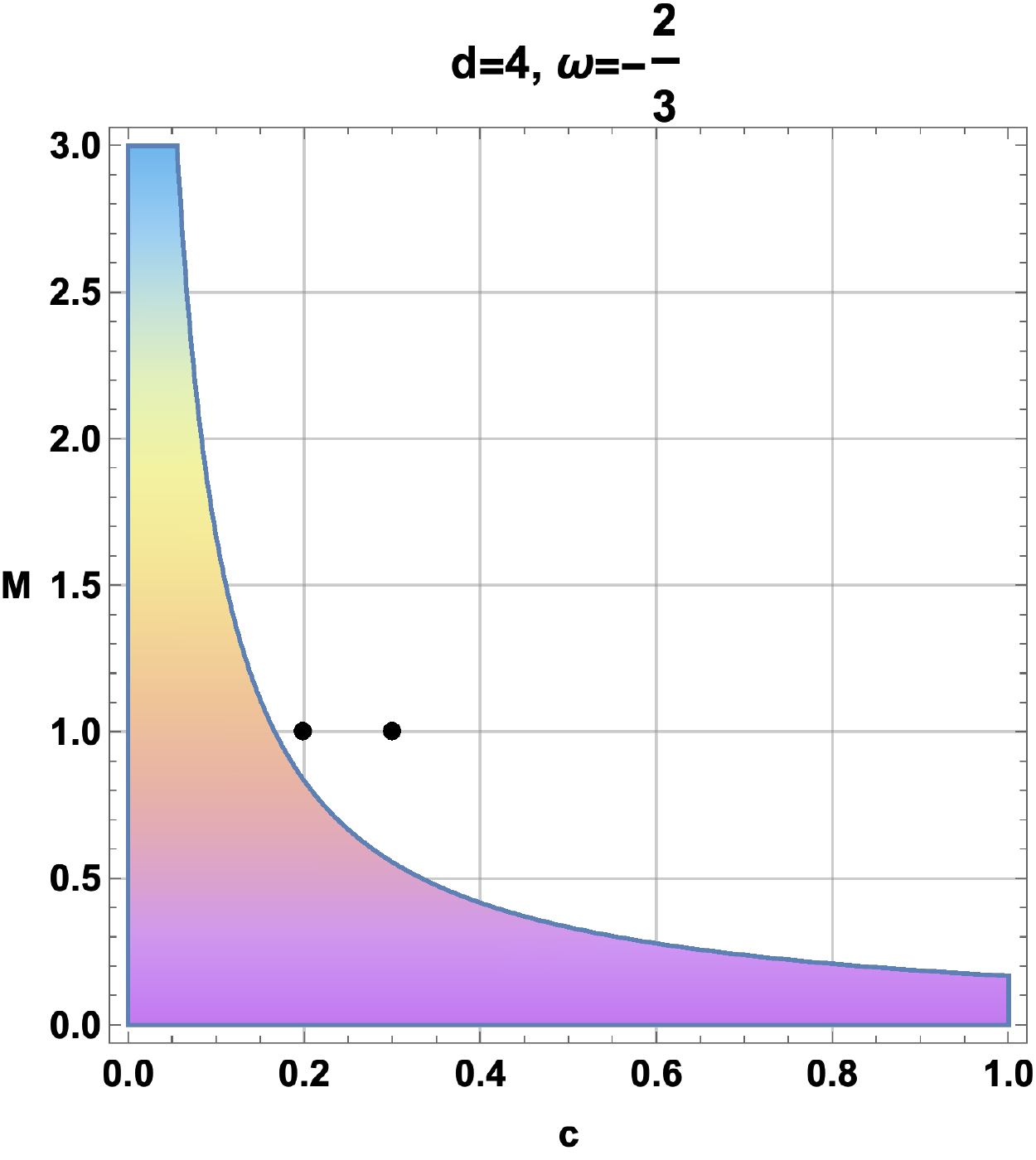} \>
			\includegraphics[scale=.56]{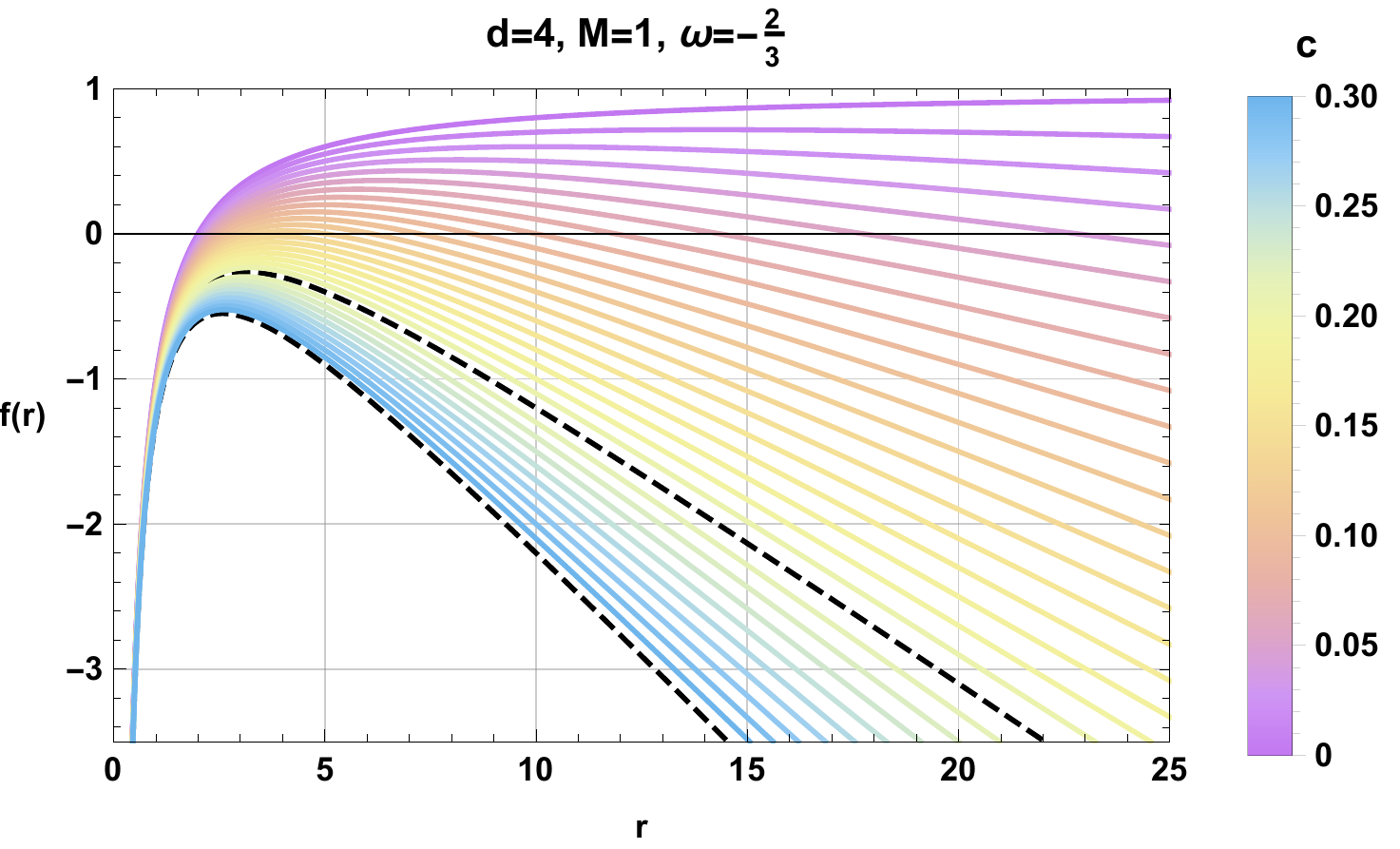} \\
		   \end{tabbing}
\caption{{\it \footnotesize Left: Region Plot presents the allowed values of $r_0$ in the diagram $M-c$.
Right: $f_\omega(r)$ in terms of $r$ with $m = 1$, and different values of $c$. The two black points  and  two dashed black lines correspond to  $c=0.2$ and $c=0.3$.}
}\label{ff21}
\end{center}
\end{figure}

To  deal with  the photon orbit, we exploit   two  impact parameters $\eta$ and $\xi$,  having  a functional form in terms of the  energy $E$, the angular momentum $L$, and the  Carter constant $\mathcal{K}$ as follows
\begin{equation}
\label{27}
 \xi=\frac{L}{E}, \hspace{1.5cm}\eta=\frac{\mathcal{K}}{E^2}.
\end{equation}
In this way,  the  effective potential and the function $\mathcal{R}(r)$  can be expressed as follows
\begin{eqnarray}
\label{28}
V_{eff} = E^2(\frac{f_\omega(r)}{r^2}(\eta+\xi^2)-1), \quad
\mathcal{R}(r) =  E^2(r^4-r^2f_\omega(r)(\eta+\xi^2)).
\end{eqnarray}
Using  \eqref{24} and  \eqref{28},  one can reveal that the   impact parameters $\eta$ and $\xi$ should satisfy
\begin{equation}
\label{30}
\eta+\xi^2=\frac{5r_0^2}{3f_\omega(r_0)+rf_\omega'(r_0)}.
\end{equation}
To analyse the relevant data,  the   Tab.\ref{t1} represents the variation of  $\eta+\xi^2$ as a function of  the dimension $d$,  the DE state parameter $\omega$ and the field intensity $c$.  It is worth noting that dimensional analysis  reveals  that $r_0$ has the dimension of the length while  $\eta+\xi^2$ has the dimension of the length square,  in $G=c=\hbar=1$ reduced units. To investigate the dimension effect on these physical quantities,  we examine  the results presented in Tab.\ref{t1}. From this table,  we observe that for fixed values of $\omega$ and $c$, $\eta+\xi^2$ decreases by  increasing   the space-time dimension $d$  in contrary to the lower  dimensions.  Fixing the dimension $d$,   $\eta+\xi^2$  increases by  increasing   $c$. Moreover, $\eta+\xi^2$  increases  generally  if one  goes from   the ($-\frac{1}{3}$)-model  to the   ($-\frac{2}{3}$)-model.  Since the calculated values of $r_0$ and $\eta+\xi^2$ shown by dots in Tab.\ref{t1} are complex  having  no physical meaning,  they are not writing in the ($-\frac{2}{3}$)-model for ceratin  values of the DE intensity $c$. Four dimensional behaviors, indeed,   can be illustrated in  Fig.\ref{ff21} for $\omega=-2/3$ with either  $c=0.2$ or $ c=0.3$.
\begin{table}[ht!]
\begin{center}
\scalebox{0.65}{\begin{tabular}{|c|c|c|c|c|c|c|c|c|l|c|c|c|c|c|}
\hline
\multicolumn{1}{|l|}{\multirow{3}{*}{}} & \multicolumn{2}{c|}{\multirow{2}{*}{$c=0$}} & \multicolumn{6}{c|}{$\omega=-\frac{1}{3}$}                                                 & \multicolumn{6}{c|}{$\omega=-\frac{2}{3}$}                                                                    \\ \cline{4-15}
\multicolumn{1}{|l|}{}                  & \multicolumn{2}{c|}{}                       & \multicolumn{2}{c|}{$c=0.1$} & \multicolumn{2}{c|}{$c=0.2$} & \multicolumn{2}{c|}{$c=0.3$} & \multicolumn{2}{c|}{$c=0.1$}                    & \multicolumn{2}{c|}{$c=0.2$} & \multicolumn{2}{c|}{$c=0.3$} \\ \cline{2-15}
\multicolumn{1}{|l|}{}                  & $r_0$          & $\eta+\xi^2$         & $r_0$  & $\eta+\xi^2$  & $r_0$  & $\eta+\xi^2$  & $r_0$  & $\eta+\xi^2$  & \multicolumn{1}{c|}{$r_0$} & $\eta+\xi^2$ & $r_0$  & $\eta+\xi^2$  & $r_0$  & $\eta+\xi^2$  \\ \hline
$d=4$                                   & 3              & 27                         & 3.333  & 37.037              & 3.750  & 52.734              & 4.285  & 78.717              & 3.675                      & 152.982            & ...    & ...                 & ...    & ...                 \\ \hline
$d=5$                                   & 1.302          & 3.395                      & 1.379  & 4.021               & 1.463  & 4.775               & 1.555  & 5.683               & 1.360                      & 4.427              & 1.429  & 6.173               & 1.519  & 9.789               \\ \hline
$d=6$                                   & 1.060          & 1.875                      & 1.117  & 2.184               & 1.177  & 2.536               & 1.240  & 2.932               & 1.093                      & 2.248              & 1.130  & 2.771               & 1.174  & 3.548               \\ \hline
$d=7$                                   & 0.993          & 1.479                      & 1.044  & 1.715               & 1.098  & 1.973               & 1.153  & 2.249               & 1.019                      & 1.732              & 1.050  & 2.067               & 1.085  & 2.525               \\ \hline
$d=8$                                   & 0.976          & 1.334                      & 1.025  & 1.539               & 1.075  & 1.754               & 1.125  & 1.976               & 1.001                      & 1.546              & 1.028  & 1.818               & 1.061  & 2.175               \\ \hline
$d=9$                                   & 0.979          & 1.279                      & 1.026  & 1.464               & 1.073  & 1.651               & 1.119  & 1.838               & 1.003                      & 1.472              & 1.030  & 1.714               & 1.062  & 2.022               \\ \hline
$d=10$                                  & 0.993          & 1.267                      & 1.036  & 1.434               & 1.079  & 1.598               & 1.120  & 1.757               & 1.015                      & 1.449              & 1.042  & 1.672               & 1.073  & 1.947               \\ \hline
$d=11$                                  & 1.011          & 1.278                      & 1.050  & 1.427               & 1.089  & 1.570               & 1.125  & 1.706               & 1.033                      & 1.451              & 1.059  & 1.658               & 1.088  & 1.908               \\ \hline
\end{tabular}
}
\caption{{\it \footnotesize  $r_0$ and $\eta+\xi^2$  in higher dimensional space-time with DE.}}
\label{t1}
\end{center}
\end{table}

It follows  from this figure  (dashed lines in left panel)  that the function $f_\omega(r)$ becomes strictly negative avoiding the formation of black hole event horizon.
It has been also observed that these non physical values are located outside the region of the values of  positive $r_0$ in the diagram $M-c$ (black dots in right panel). Moreover,  three different cases, i.e. double-horizons, single-horizon and no-horizon are depicted.

\begin{figure}[!ht]
		\centering
		\hspace{-0.5cm}
		 \includegraphics[scale=.4]{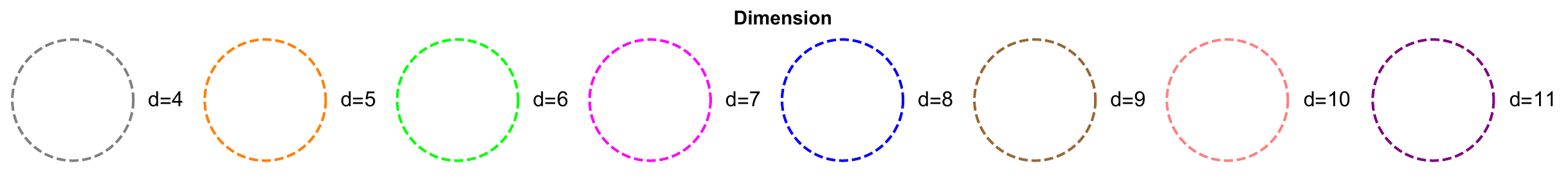}
			\begin{tabbing}
			\hspace{5.2cm}\= \hspace{5.2cm}\=\kill
			\includegraphics[scale=.39]{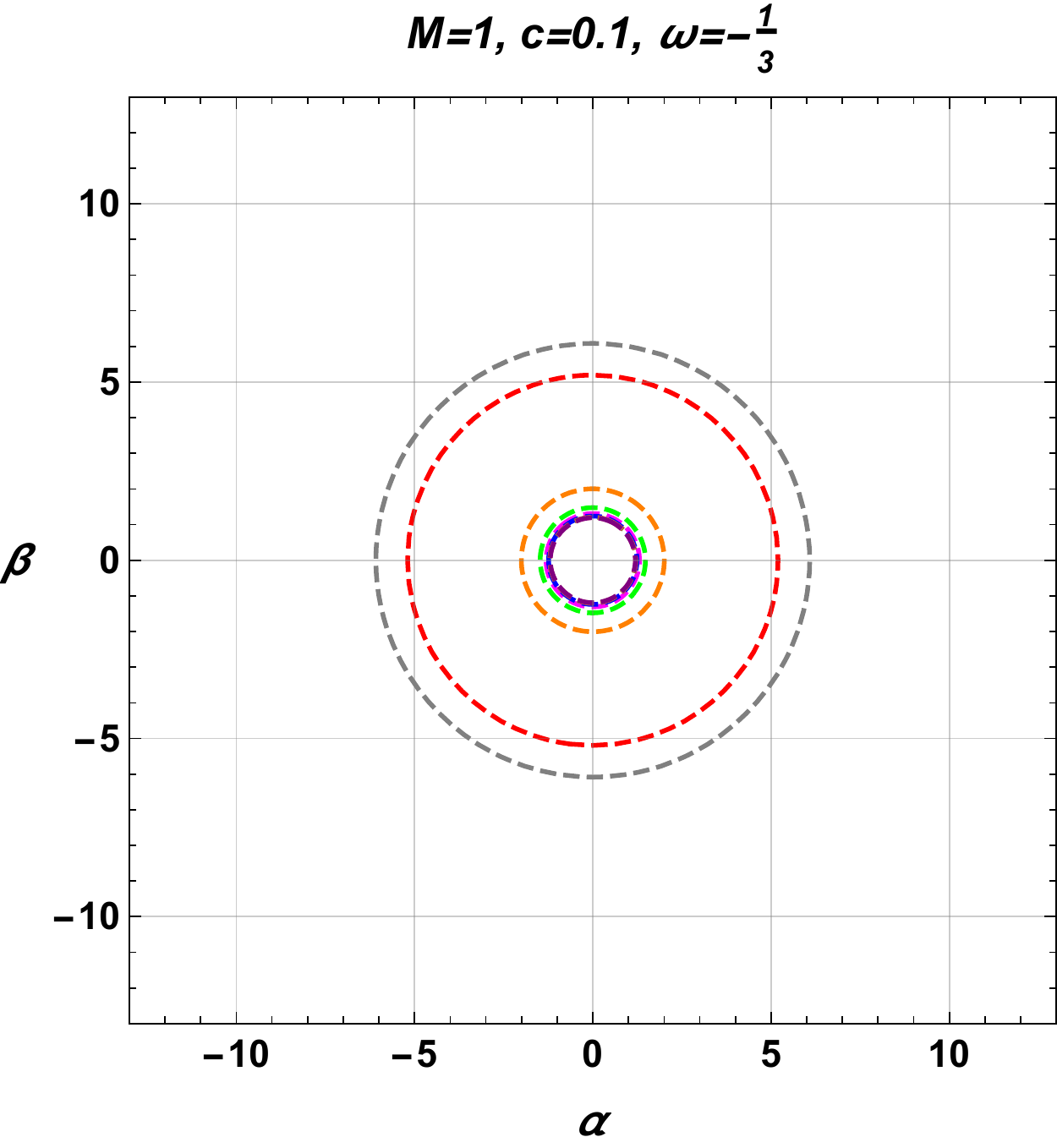} \>
			\includegraphics[scale=.39]{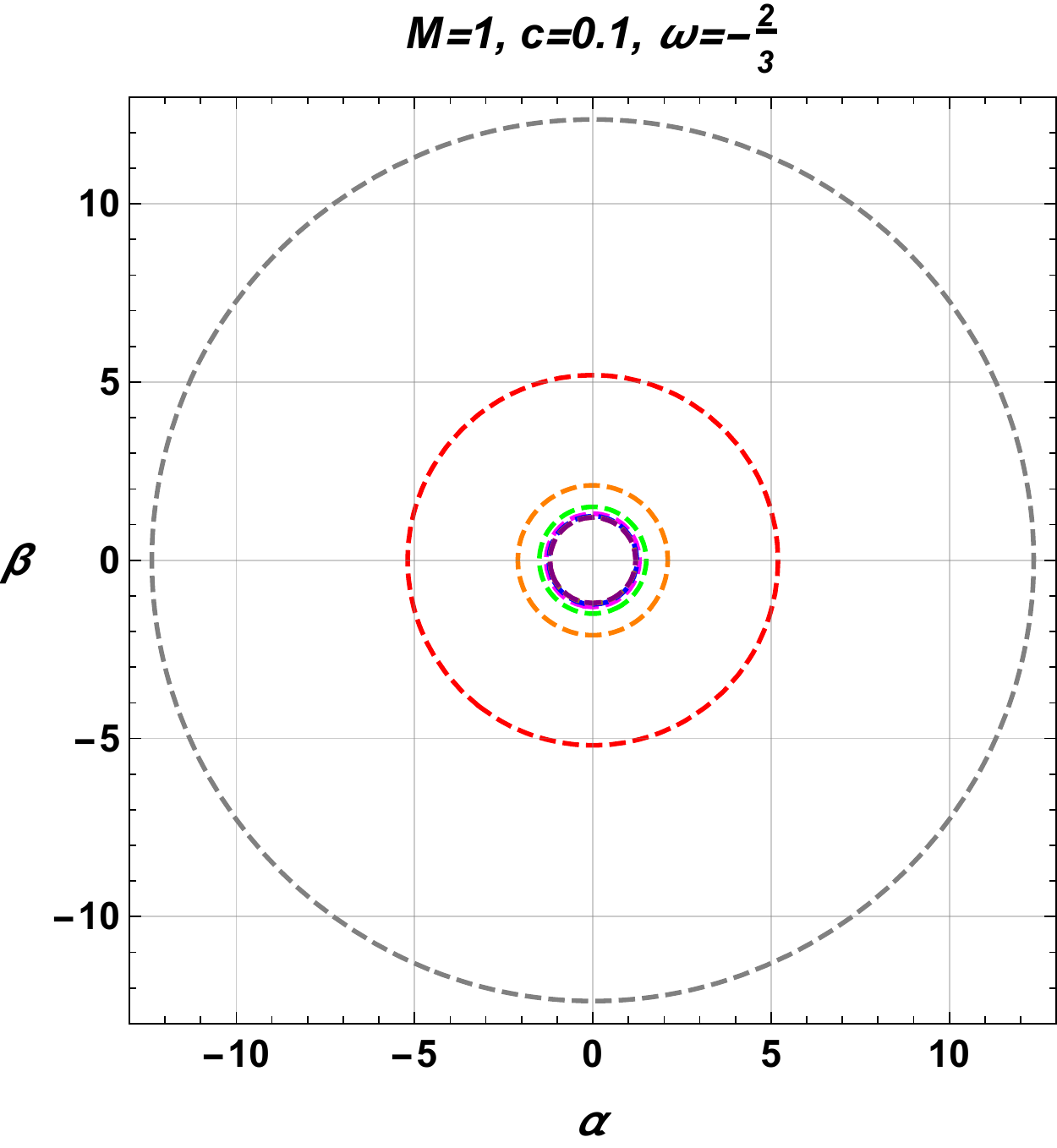} \>
			\includegraphics[scale=.39]{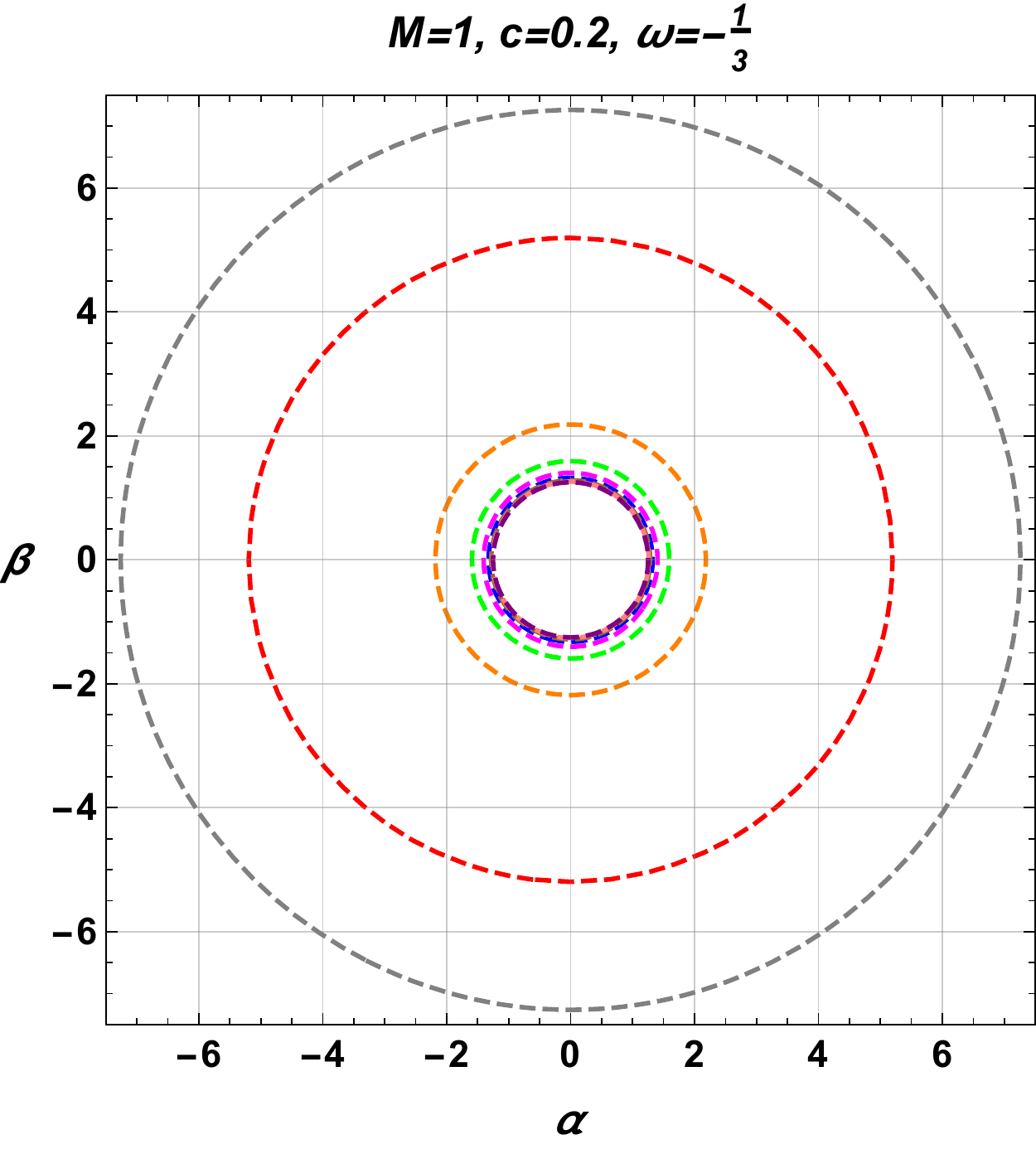} \\
			\includegraphics[scale=.39]{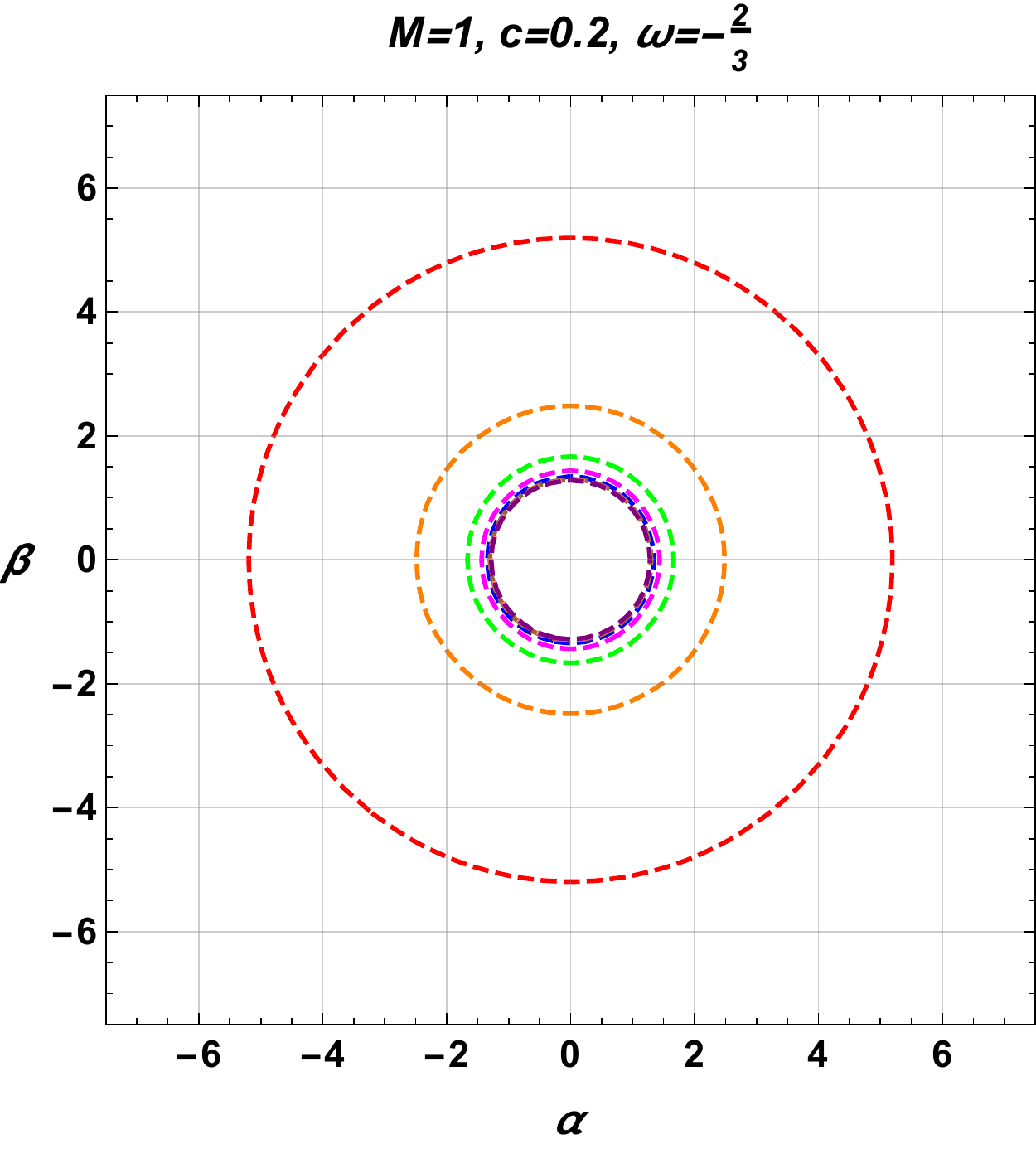} \>
			\includegraphics[scale=.39]{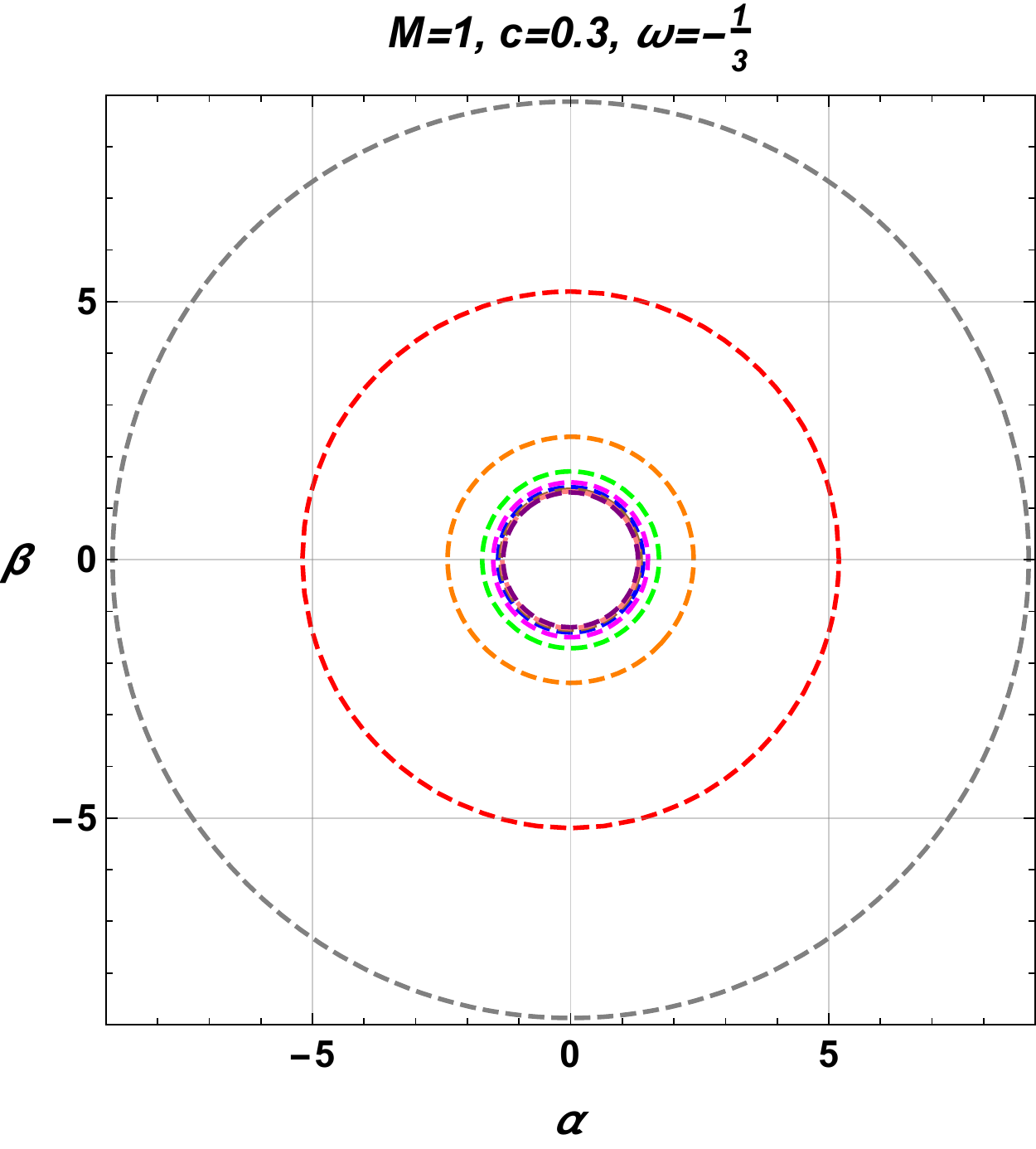} \>
			\includegraphics[scale=.39]{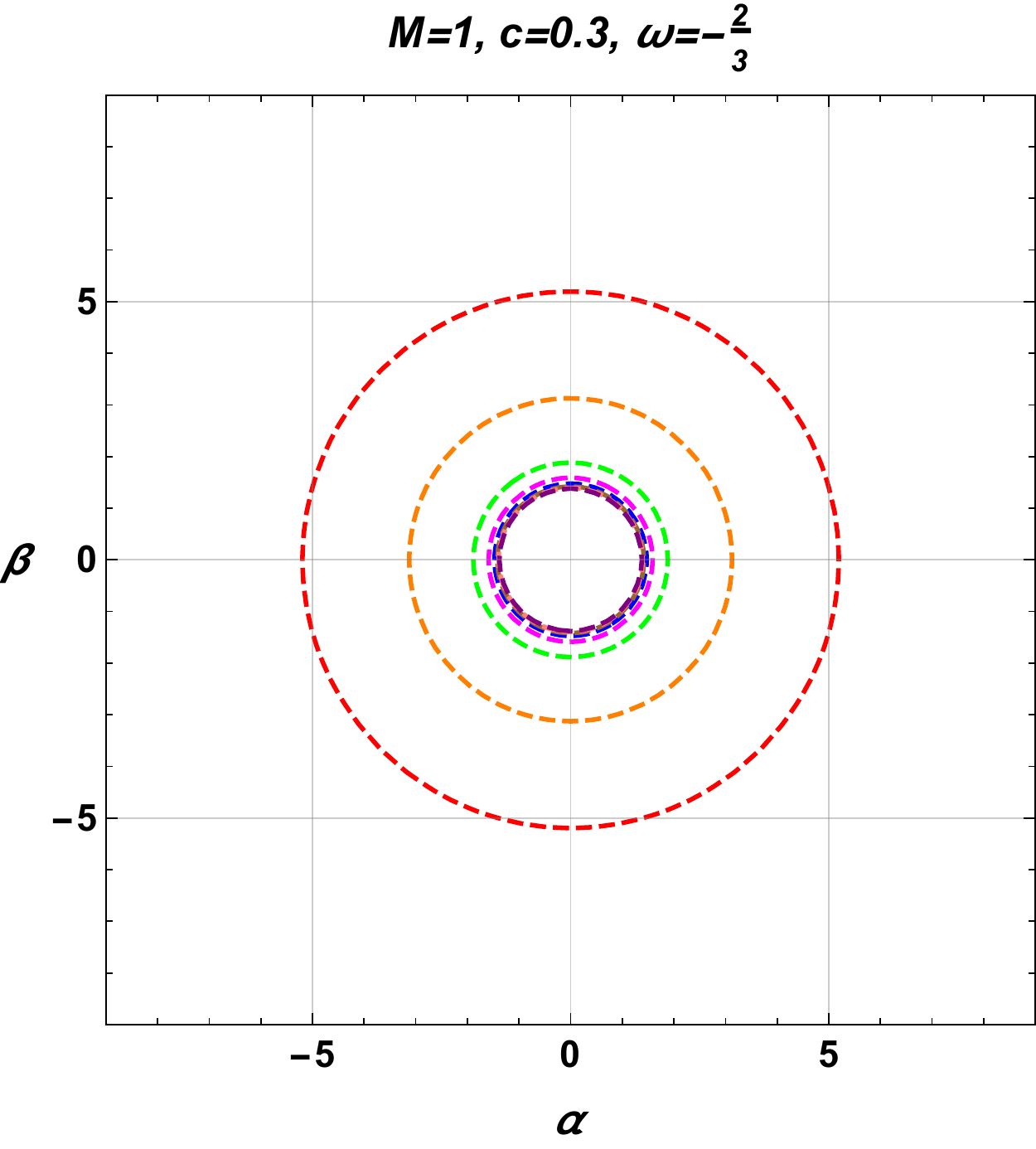} \\
			
				        \end{tabbing}
	
\caption{\it \footnotesize  Black hole shadow in the celestial plane ($\alpha-\beta$ plane) for both models   $(-\frac{1}{3})$ and $(-\frac{2}{3})$ within the dimension $d$ and the field intensity $c$.  In the all panels,  the red circle corresponds to the  Schwarzschild black hole shadow without DE.}
\label{f2}
\end{figure}

  To properly visualize the shadow on the observer's frame,   one should use  the celestial coordinates $\alpha$ and $\beta$ reported in \cite{vazquez2003strong}.  Following to  \cite{chandrasekhar1985mathematical}, the celestial coordinates $\alpha$ and $\beta$ have been taken  as follows
  \begin{eqnarray}
\label{31}
\alpha  =   \lim_{r_0 \to \infty} (\frac{r_0{P^{(\theta_{d-2})}}}{P^{(t)}}),  \quad
\beta_i   =   \lim_{r_0 \to \infty} (\frac{r_0{P^{(\theta_{i})}}}{P^{(t)}}), \hspace{1cm} i=1, \ldots, d-3,
\end{eqnarray}
where $r_0$ is  the distance between the black hole and a far distant observer, and $\left[P^{(t)},P^{(\theta_{d-2})},P^{(\theta_{i})}\right]$  are  the vi-tetrad component of momentum.
Placing the observer  on  the equatorial hyperplane,  these  equations are reduced to
\begin{equation}
\label{35}
\alpha=-\xi, \hspace{1cm} \beta=\pm\sqrt\eta.
\end{equation}
In this way,  equation \eqref{30}  can be  rewritten as
\begin{equation}
\label{36}
\eta+\xi^2=\alpha^2+\beta^2.
\end{equation}
It is worth noting that,  in the absence of DE, we recover  the   Schwarzschild  black hole   result \cite{singhshadow}. To inspect the  DE effect on   the shadow  geometric  circular shape, we plot, in  Fig.\ref{f2}, the  associated size behavior  of   $(-\frac{1}{3})$ and $(-\frac{2}{3})$-models in arbitrary  dimension $d$ as a function of $c$.
It follows  from this figure that DE  can be considered as  a size shadow  parameter. In particular, the associated size increases by  increasing  the field intensity  $c$. Similar behaviors are observed with  the  DE state parameter.  Switching  from  $(-\frac{1}{3})$-model to $(-\frac{2}{3})$-one, for  a fixed  $c$ value,  this brings  an increasing  size circular geometry.  Concretely, the present  study reveals that DE  leads to a  violation
of  some bounds  suggesting that the Schwarzschild solution  is the biggest  of all black holes for given masses \cite{Hod:2017xkz,Lu:2019zxb}.\\
However,   the increasing  of   the  space-time dimension $d$  reduces the  shadow circular  size.  For dimensions $d>11$,   such a size  remains constant  allowing one to consider  $d=11$ as  a critical one  for the shadow of the Schwarschild-Tanglerlini black hole  with DE.  It is worth noting that such  a dimension,   associated with a known theory called M-theory,   has been approached in connection with dark sector from string  axion fields \cite{boby}. It  should be interesting  to  unveil certain links with M-theory in future works by focusing on such   non-trivial stringy  fields.

\begin{figure}[ht!]
\begin{center}
\includegraphics[width=16cm, height=5.cm]{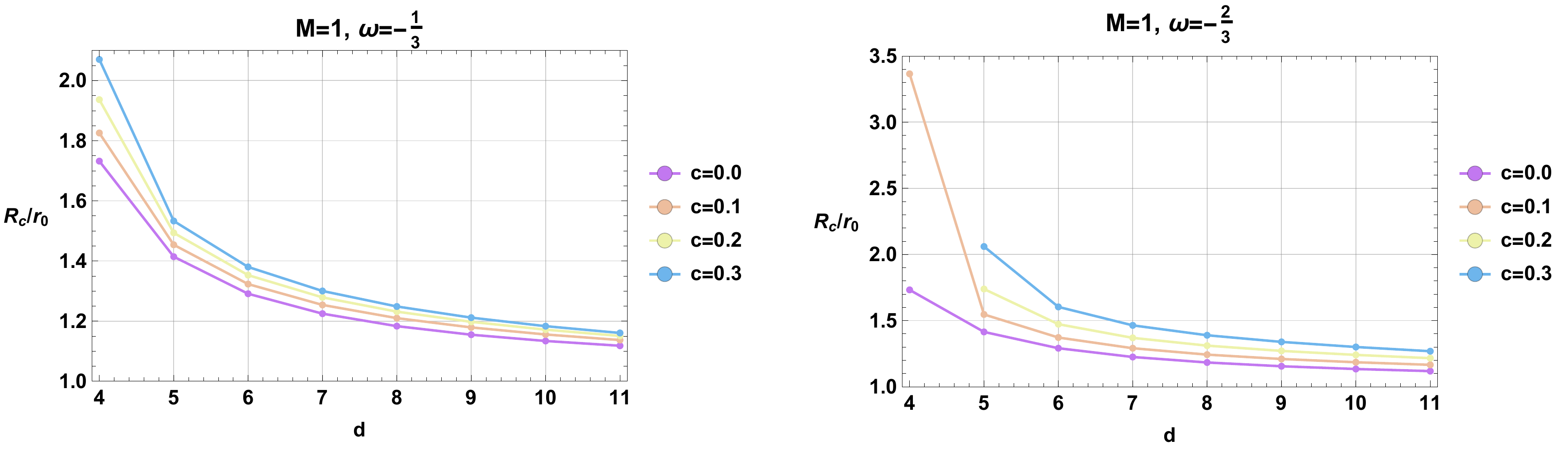}
\end{center}
\caption{\it \footnotesize Variation of shadow radius $R_c$ over photon sphere radius $r_0$ as a function of dimension $d$, associated with  $(-\frac{1}{3})$,  $(-\frac{2}{3})$-models  for  different values of the intensity $c$.}
\label{f3}
\end{figure}

 More inspections,  on  the photon behavior  gravitating  around the black hole at the distance of the photon sphere $r_0$ allow one to consider the ratio $ \frac{R_c}{r_0}$ as a  function  of the dimension $d$. This is illustrated in    Fig.\ref{f3}.

 It has been observed that for  $d>6$ the radius of the photon sphere  and   the radius of the shadow circle  are  almost the same. Concretely,  they are approximately confused  for  $(-\frac{1}{3})$ and $(-\frac{2}{3})$-models for  different  values  $c$.  However,   $d\leq 6$,  the radius $R_c$ is larger  with respect to $r_0$.

\subsection{Energy emission rate}
\begin{figure}[htb]
		\begin{center}
		\centering
		\begin{tabbing}
			\centering
			\hspace{8.9cm}\=\kill
			\includegraphics[scale=.32]{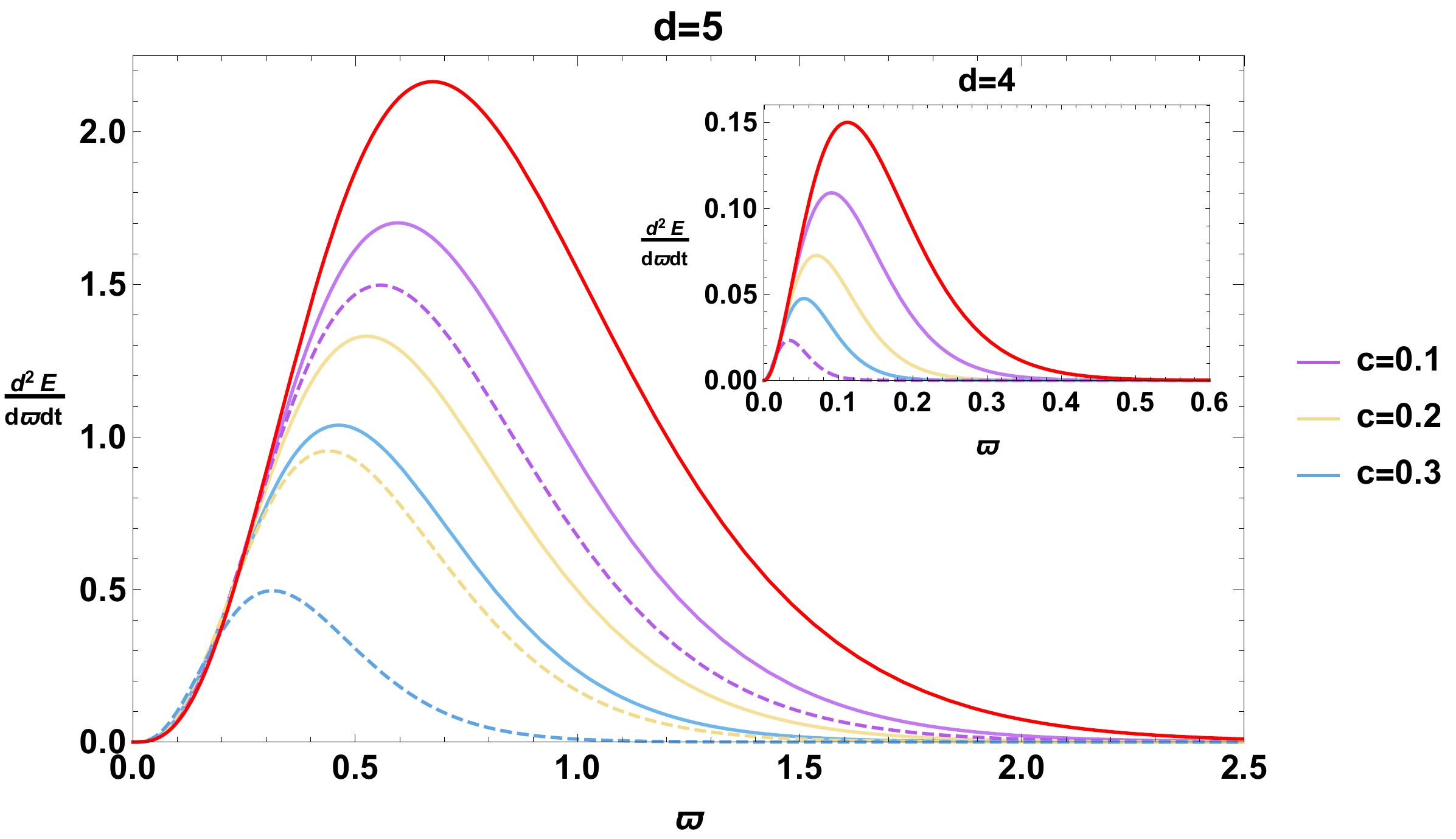} \>
			\includegraphics[scale=.32]{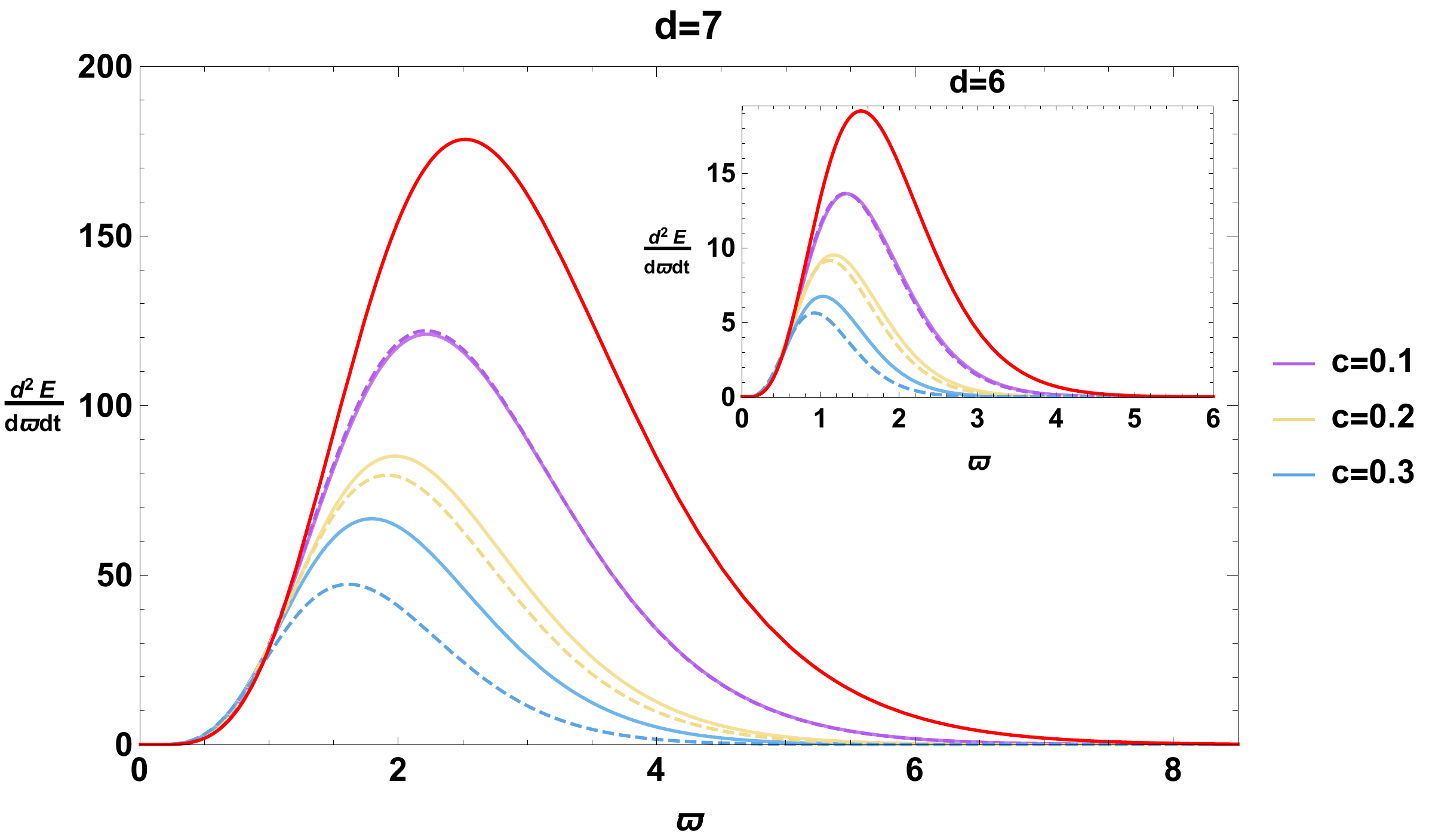} \\
			\includegraphics[scale=.32]{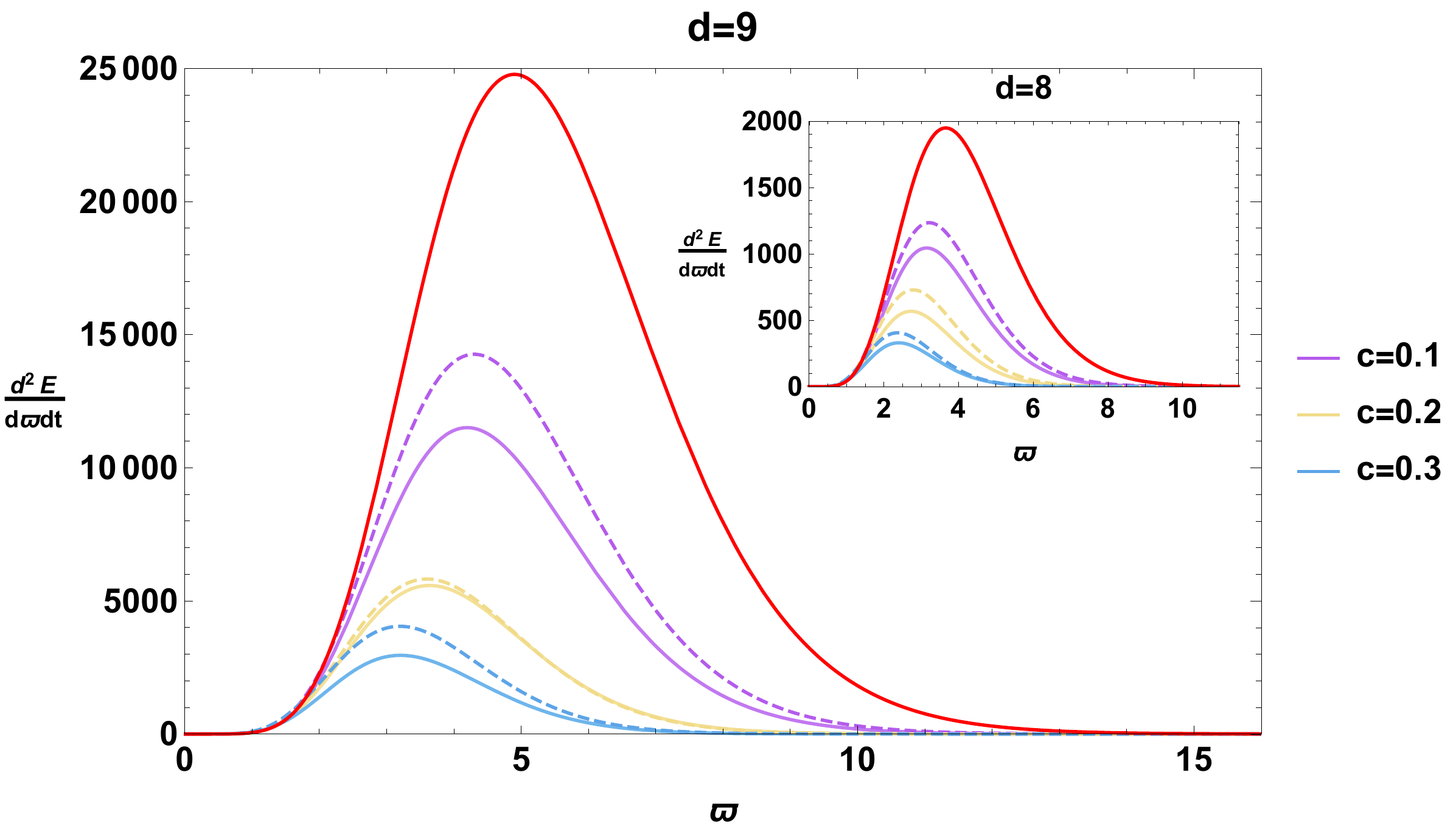} \>
			\includegraphics[scale=.32]{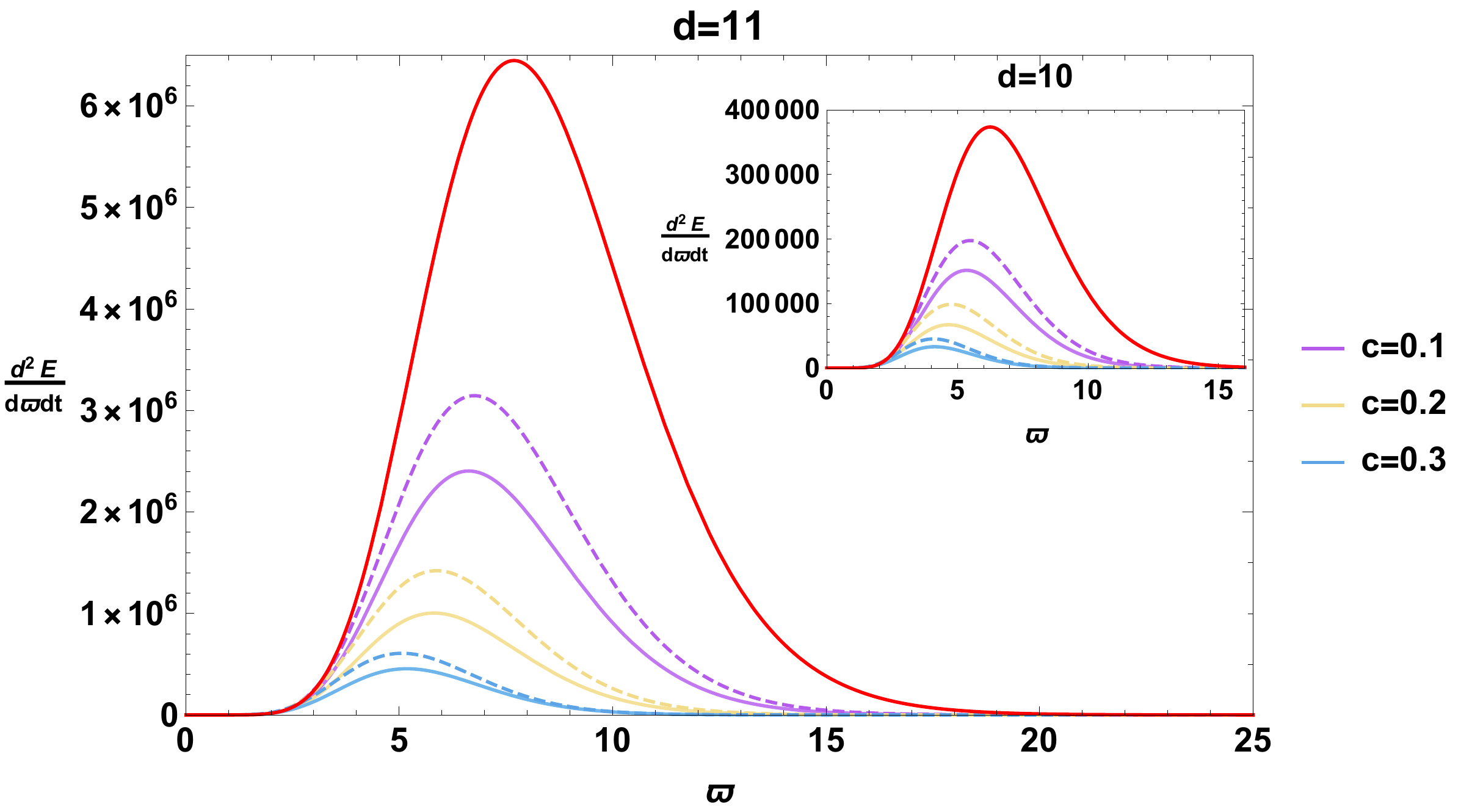} \\
			\end{tabbing}
\caption{{\it \footnotesize  Energy emission rate for different dimensions $d$ and  different values of DE the intensity $c$ where the red curve represents the case without DE. The straight lines represent the value $\omega=-1/3$ and the dashed ones are associated with  $\omega=-2/3$.}
}
\label{er1}
\end{center}
\end{figure}

   It is known that, inside the black holes,  quantum fluctuations create and annihilate a large number of particle pairs near the horizon. In this way, the positive energy particles escape through tunneling from the black hole, inside region where the Hawking radiation occurs.  This process is known as the  Hawking radiation  causing  the black hole to evaporate in a certain period of time. Here,  we study  the associated energy emission rate. In this case,  for a far distant observer the high energy absorption cross section approaches to  the black hole shadow. The absorption cross section of the black hole oscillates to a limiting constant value $\sigma_{lim}$ at very high energy.  It  turns out that the limiting constant value,  being  approximately equal to the area of photon sphere, can be expressed as
\begin{equation}
\frac{d^2 E(\varpi)}{d\varpi dt}=\frac{2\pi^2 \sigma_{lim}}{e^{\frac{\varpi}{T_{{out}}}}-1}\varpi^{(d-1)},
\end{equation}
 where $\varpi$   indicates  the emission frequency \cite{Wei:2013kza}. It is noted that  $T_{out}$ is the  Hawking temperature for  the Schwarzschild-Tangherlini with DE \cite{belhaj_dark_2019}.    Such a temperature  reads as
\begin{equation}
T_{{out}}= \frac{1}{4 \pi }\left(\frac{d-3}{ r}+c (d-1) \omega  r^{-d (\omega +1)+\omega +2}\right).
\end{equation}
According to \cite{m27,m28,decanini_fine_2011}, for a  higher-dimensional space-time,   $\sigma_{lim}$ can be given by
\begin{equation}
\label{71}
\sigma_{lim}\approx \frac{\pi^{(d-2)/2}R_c^{d-2}}{\Gamma(\frac{d}{2})}.
\end{equation}
Using  \eqref{71}, we get the expression of the Schwarzschild-Tangherlini black hole energy emission rate in the presence of DE  in higher-dimensional space-time as
\begin{equation}
\label{72}
\frac{d^2 E(\varpi)}{d\varpi dt}=\frac{2\pi^{(d+2)/2}(\varpi R_c)^{d-2}}{(e^{\frac{\varpi}{T_{{out}}}}-1)\Gamma(\frac{d}{2})}\varpi.
\end{equation}
 The energy emission rate is illustrated in Fig.\ref{er1} as a function of $\varpi$ for different space-time  dimensions and values of the DE intensity $c$.
It is observed from Fig.\ref{er1} that, when DE is present,  the energy emission rate is lower meaning that the black hole evaporation process is slow. Besides, we obtain an even slower radiation process by decreasing (increasing) the state parameter $\omega$ (the intensity $c$). However, increasing the dimension of the black hole implies a fast emission of particles. This shows that the evaporation of a higher dimensional black hole is fast compared to the one living in four dimensions. Furthermore,  we  can notice a special behavior for certain  particular dimensions.   For instance, the energy emission rate for the cases $d=6$ and $d=7$ for $\omega=-1/3$ and $\omega=-2/3$ models matches perfectly. This implies that some of the dimensions may show a resistance regarding the change of the state parameter.  Taking into account of  the studied models,  the variation of the energy emission rate with respect to the space-time dimension reveals an intrigued behavior. Comparing the solid and dashed lines of each panel, one can notice that for $4\leqslant d \leqslant 7$ the emission associated with the  $\omega=-\frac{1}{3}$ is more important than $\omega=-\frac{2}{3}$. However,  for  $8\leqslant d \leqslant 11$ the situation is inverted.
\vspace{7cm}

\section{Deflection angle behavior  of QBH in arbitrary dimensions}
\label{secdeflection}
\begin{figure}[!ht]
		\begin{center}
		\centering
			\begin{tabbing}
			\centering
			\hspace{9.3cm}\=\kill
			\includegraphics[scale=.5]{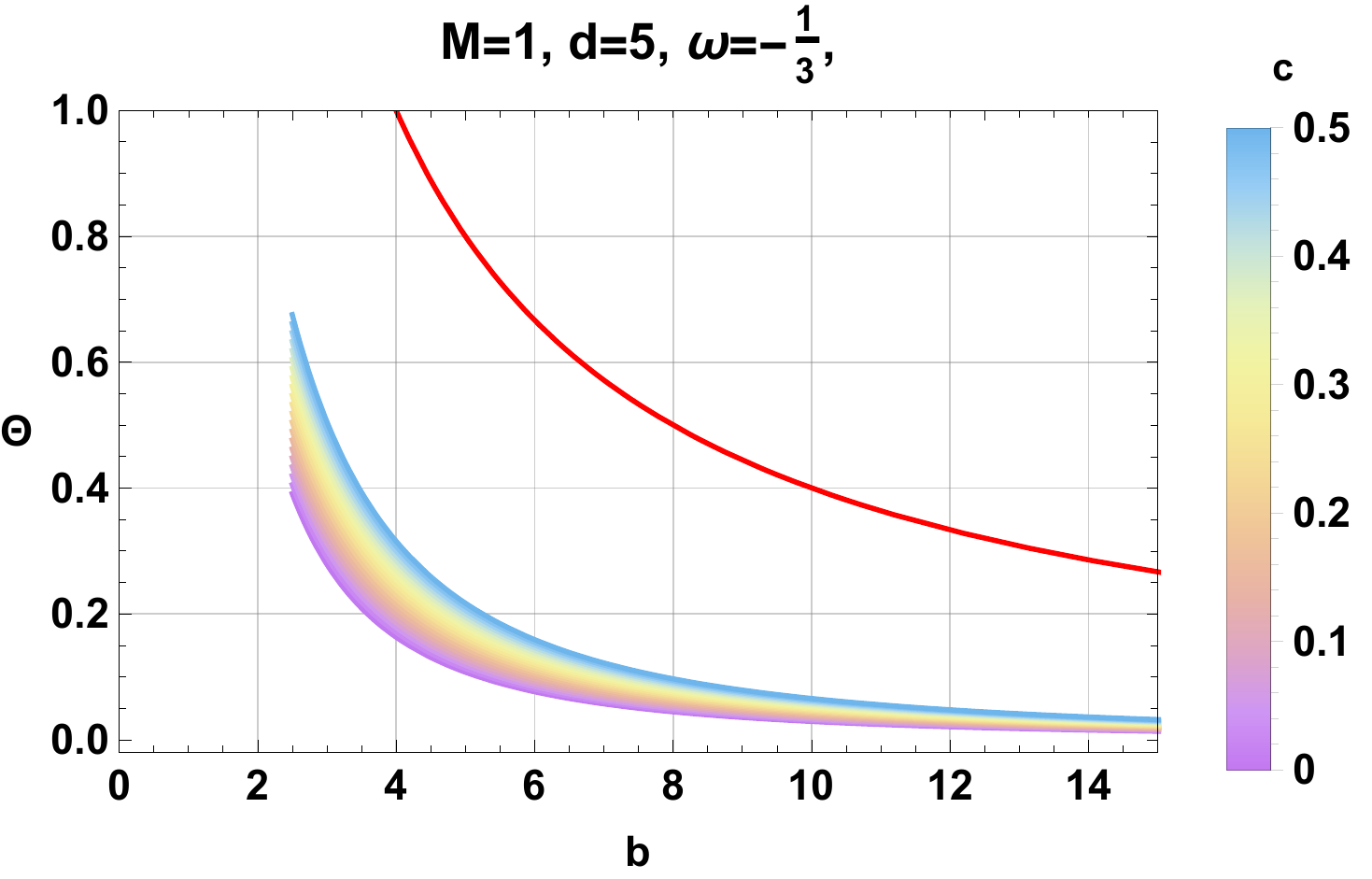} \>
			\includegraphics[scale=.5]{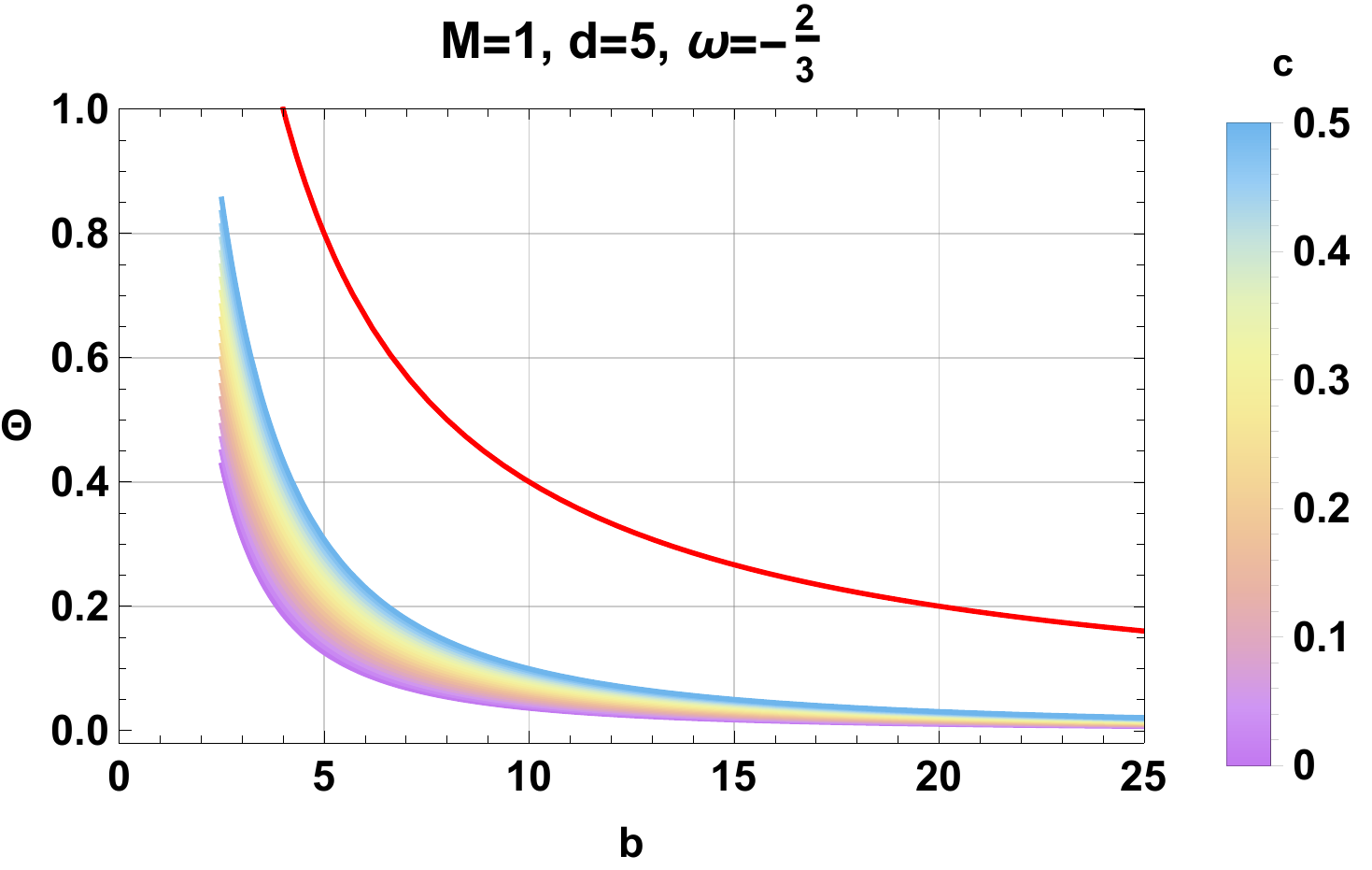} \\
			\includegraphics[scale=.5]{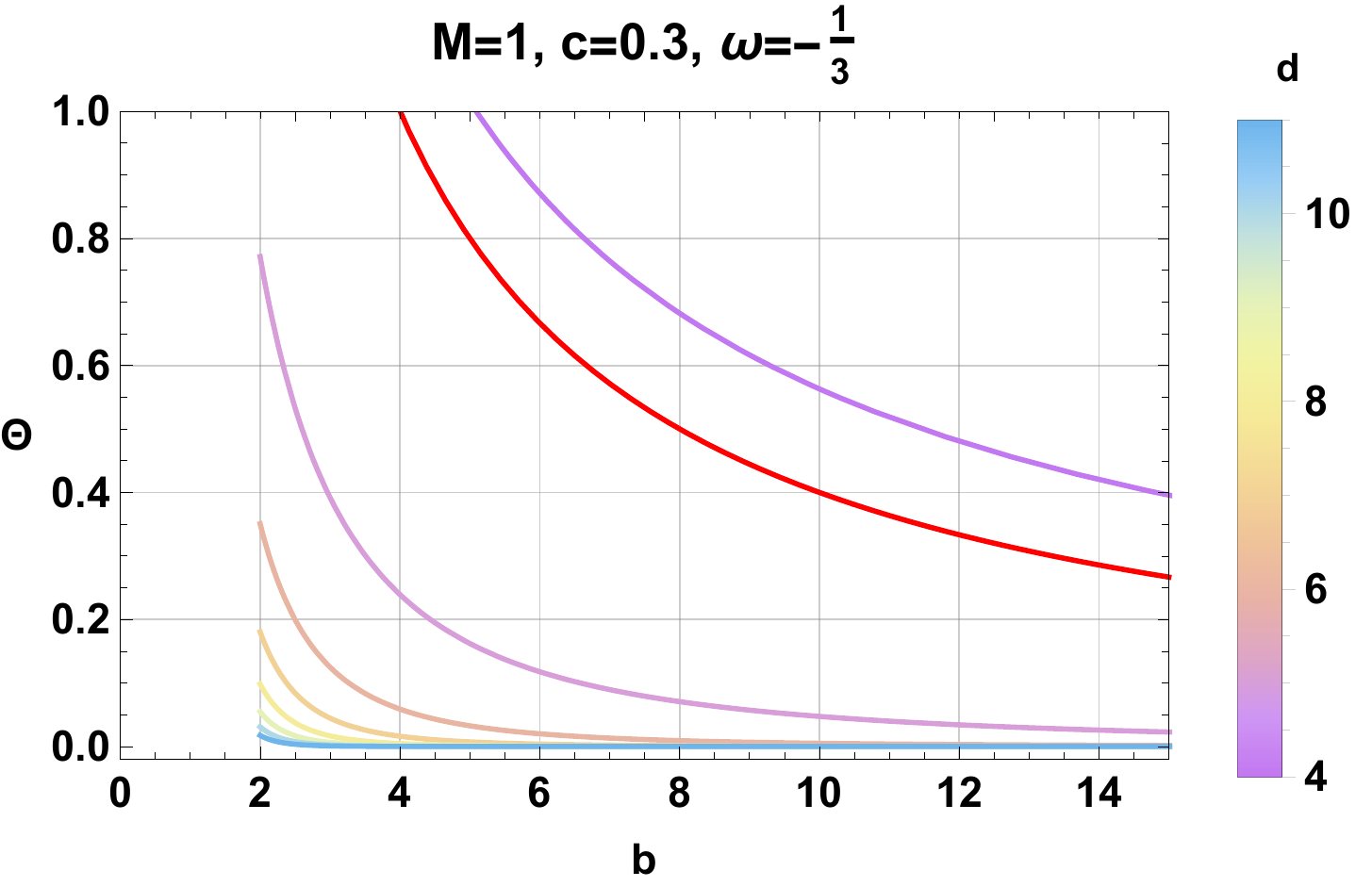} \>
			\includegraphics[scale=.5]{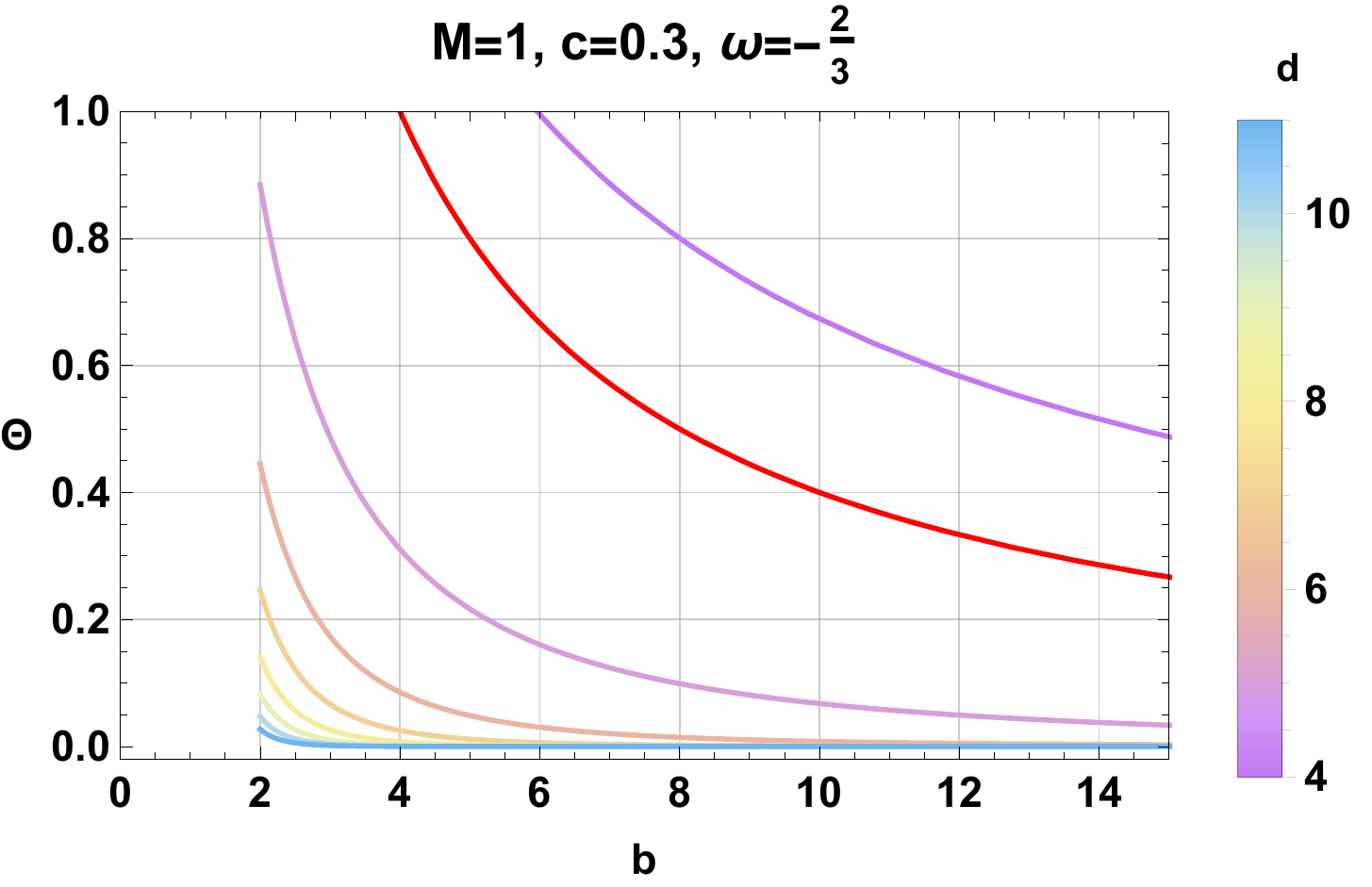} \\
		   \end{tabbing}
\caption{{\it \footnotesize Variation of the deflection angle  as a function of the  parameter $b$ in different dimension $d$, $c$ and  for two values of $\omega$.  In the all panels,  the red curve corresponds to deflection of Schwarzschild black hole without DE.
}}\label{f4}
\end{center}
\end{figure}

In this section, we study the behaviors of  the  deflection angle of  quintessential Shwarzschild-Tanglerlini black holes  by analysing  the effect of various  parameters including  the space-time dimension $d$ and  DE. It is recalled that   such an angle     can be computed  from the relation
\begin{equation}
\label{38}
\Theta=-\iint_{S_\infty}{K dS},
\end{equation}
where $K$ denotes  the Gaussian curvature and where  $dS$  is the surface of the associated  optical metric \cite{Gibbons:2008rj}.   It has been shown that this equation  can be obtained by combining such a  optical metric and the  Gauss-Bonnet theorem.  It is  noted   for a space denoted ($D_R$,$\chi$,$g$) where  $D_R$ is the  relevant region with  a geometrical size  $R$,  $\chi$ is  the associated  Euler characteristic and $g$ is   the corresponding  Riemannian metric, the   Gauss-Bonnet theorem  stipulates
 \begin{equation}
\label{50}
\iint_{D_R}{KdS}+\oint_{\partial D_R}{kdt}+\sum_i\eta_i=2\pi\chi(D_R).
\end{equation}
Here, $k$ is  the geodesic curvature given by  $k=\bar{g}(\nabla_{\dot{\alpha}}\dot{\alpha},\ddot{\alpha})$ where $\ddot{\alpha}$ is the unit acceleration vector.  For $R$ goes to $ \infty$,  the jump angles $ \alpha_s$(source) and $\alpha_O$(observer) become $\alpha_s=\alpha_o=\frac{\pi}{2}$. The source and the  observer interior angles are  $\eta_s=\pi-\alpha_s$ and $\eta_0=\pi-\alpha_0$. For $\chi(D_R)=1$  associated with a   non-singular behavior,  the Gauss-Bonnet theorem reduces to
\begin{equation}
\label{51}
\iint_{D_R}{KdS}+\oint_{\partial D_R}{kdt}+\eta=2\pi\chi(D_R),
\end{equation}
where $\eta=\eta_O+\eta_s=\pi$.  To obtain the relevant quantities
  including the Gaussian curvature, one should consider   the equatorial hyperplane  $\theta_i(i=1,\ldots,d-3)=\frac{\pi}{2}$.
 Using the notation $d\theta_{d-2}=d\phi^2$,  the metric of the quintessential  Schwarschild-Tanglerlini black holes  given in  \eqref{2} becomes
 \begin{equation}
\label{41}
ds^2=f_\omega(r)dt^2+f_{\omega}(r)^{-1}dr^2+r^2d\phi^2,
\end{equation}
where now $ds^2$  denotes  the optical metric. For null geodesics $ds^2=0$, one gets  the  optical metric tensor
\begin{equation}
\label{42}
dt^2=\frac{1}{f_{\omega}(r)^2}dr^2+\frac{r^2}{f_{\omega}(r)}d\phi^2.
\end{equation}
In this way, the Gaussian curvature  in the presence of   DE can be obtained  from the equation
\begin{equation}
\label{43}
 K=\frac{{\mathbf{R}}}{2},
\end{equation}
 where  ${\mathbf{R}}$ is the associated   Ricci scalar.  To get such quantities,   the Christoffel symbols are needed. Indeed, the  non-zero Christoffel symbols are given
\begin{eqnarray}
\label{54}
\Gamma^r_{rr} & = & -\frac{f_{\omega}'(r) }{f_{\omega}(r)},\\
\label{55}
\Gamma^r_{\phi\phi} & = & \frac{1}{2}r^2f_{\omega}'(r)-rf_{\omega}(r),\\
\label{56}
\Gamma^\phi_{r\phi} & = & \frac{1}{r}-\frac{f_{\omega}'(r) }{2f_{\omega}(r)},
\end{eqnarray}
where  one has used $f_{\omega}'(r)=\frac{\partial f_{\omega}(r)}{\partial r}$. It is noted that  the  Ricci scalar  for the optical metric  reads as
\begin{equation}
\label{44}
{\mathbf{R}}= -\frac{1}{2}\Big(\frac{\partial f_\omega(r)}{\partial r}\Big)^2+f_\omega(r)\frac{\partial^2f_\omega(r)}{\partial r^2}.
\end{equation}
The calculation shows that
\begin{equation}
\label{45}
\begin{split}
f_\omega(r)\frac{\partial^2f_\omega(r)}{\partial r^2}=&-\bigg(1-\frac{\mu}{r^{d-3}}-\frac{c}{r^{\omega(d-1)+d-3}}\bigg)\bigg(\frac{\mu(d^2-5d+6)}{r^{d-1}}\\
&+c\Big(\frac{d^2(\omega^2+2\omega+1)}{r^{\omega(d-1)+d-1}}-\frac{d(2\omega^2+7\omega+5)+(\omega^2+5\omega+6)}{r^{\omega(d-1)+d-1}}\Big)\bigg),
\end{split}
\end{equation}
together with
\begin{equation}
\label{47}
\begin{split}
\Big(\frac{\partial f_\omega(r)}{\partial r}\Big)^2=&\frac{(d-3)^2\mu^2}{r^{2d-4}}+\frac{c^2\big(d(\omega+1)-(\omega+3)\big)^2}{r^{2\omega(d-1)+2(d-2)}}\\
&+\frac{2\mu c\big((d-3)(d(\omega+1)-(\omega+3))\big)}{r^{\omega(d-1)+2(d-2)}}.
\end{split}
\end{equation}
Using\eqref{43},   we obtain
\begin{equation}
\label{48}
\begin{split}
K=&-\mu\frac{(d^2-5d+6)}{2r^{d-1}}+\mu^2\frac{(d^2-4d+3)}{4r^{2(d-2)}}\\
 &-c\Big(\frac{d^2(\omega^2+2\omega+1)-d(2\omega^2+7\omega+5)+(\omega^2+5\omega+6)}{2r^{\omega(d-1)+d-1}}\Big)\\
 &+c\mu\Big(\frac{d^2(\omega^2+\omega+1)-d(2\omega^2+3\omega+4)+(\omega^2+2\omega+3)}{2r^{\omega(d-1)+2(d-2)}}\Big)\\
 &+c^2\Big(\frac{d^2(\omega^2+2\omega+1)-d(2\omega^2+6\omega+4)+(\omega^2+4\omega+3)}{4r^{2\omega(d-1)+2(d-2)}}\Big).\\
 \end{split}
\end{equation}
For simplicity  reasons, we consider  the  Gaussian optical curvature up to the  leading orders ($\mathcal{O}$($M^2$,$c^2$)) given by
\begin{equation}
\label{49}
\begin{split}
K\approx&-\mu\frac{(d^2-5d+6)}{2r^{d-1}}+\\
 &-c\Big(\frac{d^2(\omega^2+2\omega+1)-d(2\omega^2+7\omega+5)+(\omega^2+5\omega+6)}{2r^{\omega(d-1)+d-1}}\Big)\\
 &+c\mu\Big(\frac{d^2(\omega^2+\omega+1)-d(2\omega^2+3\omega+4)+(\omega^2+2\omega+3)}{2r^{\omega(d-1)+2(d-2)}}\Big).\\
 \end{split}
 \end{equation}
To determine the deviation from the geodesic,  it is useful to use the geodesic curvature
\begin{equation}
\label{52}
k(\gamma_R)=|\nabla_{\gamma_R}\dot{\gamma_R}|,
\end{equation}
$\gamma_R$ is  a geodesic. Assuming that  $\gamma_R=r(\phi)=R=const$, the radial part of the geodesic curvature reads as
\begin{equation}
\label{53}
(\nabla_{\gamma_R}\dot{\gamma_R})^r=\dot{\gamma_R}^\phi\partial_\phi\dot{\gamma_R}^r+\Gamma^r_{\phi\phi}(\dot{\gamma_R}^\phi)^2.
\end{equation}
 According to \cite{Gibbons:2008rj}, the second term  provides
\begin{equation}
\label{61}
\oint_{\partial D_R}{kdt}=\pi+\Theta.
\end{equation}
Using the linear approach of the light  ray $(r=\frac{b}{\sin\phi})$ and equation \eqref{61},  the deflection angle
 becomes
 \begin{equation}
\label{62}
 \Theta=-\int_0^\pi{\int_\frac{b}{\sin\phi}^\infty{KdS}},
\end{equation}
where $b$ is called the  impact parameter and $dS$ is given  by
\begin{equation}
\label{63}
dS=\sqrt{det\bar{g}}\frac{dr}{f_\omega(r)}\approx rdr(1+\frac{3\mu}{2r^{d-3}}+\frac{3c}{r^{2\omega(d-1)+d-3}}).
\end{equation}
Indeed,  the deflection angle is  approximated as follows
\begin{equation}
\label{64}
\Theta\approx-\int_0^\pi{\int_\frac{b}{\sin\phi}^\infty{Krdr}}.
\end{equation}
Using the result $
\int_0^\pi{\int_\frac{b}{\sin\phi}^\infty{r^{-n}}}=\frac{\sqrt{\pi } n b^{1-n} \Gamma \left(\frac{n}{2}\right)}{(n-1) \Gamma \left(\frac{n+1}{2}\right)}$ and the expression of $K$ given in \eqref{49}, we get
\begin{equation}
\label{67}
\begin{split}
\Theta\approx&\frac{c\sqrt{\pi } ( \omega(d-1) +d-2)  \Gamma \left(\frac{1}{2} ( \omega(d-1) +d-2)\right)b^{\omega(1-d) +3-d}}{2 \Gamma \left(\frac{1}{2} (d-1) (\omega +1)\right)}\\
&-\Bigg(\frac{ c\mu\sqrt{\pi } (d-1)   \left(\omega ^2(d-1)+ \omega(d-2) +d-3\right)}{2 ( \omega(d-1) + 2d-6)}\\
&\times\frac{ \Gamma \left(\frac{1}{2} (\omega(d-1) +2d-5)\right) b^{-d (\omega +2)+\omega +6}}{ \Gamma \left( \frac{\omega}{2} (d-1) +d-2\right)}\Bigg)+ \frac{\mu\sqrt{\pi }    \Gamma \left(\frac{d}{2}\right)b^{3-d}}{\Gamma \left(\frac{d-1}{2}\right)}+\mathcal{O}(M^2,c^2).
\end{split}
\end{equation}
Applying a  power series expansion, we obtain  the deflection angle of  the  quintessential Schwarschild-Tanglerlini black holes
\begin{equation}
\label{68}
\Theta\approx\frac{\sqrt{\pi } \mu  b^{3-d} \Gamma \left(\frac{d}{2}\right)}{\Gamma \left(\frac{d-1}{2}\right)}+\frac{c\omega(d-1)\sqrt{\pi }}{4} (A-B)+\mathcal{O}(M^2,c^2,\omega^2),
\end{equation}
where  the involved terms  $A$ and   $B$   take the following form
\begin{eqnarray}
A & = &\frac{ \mu   \Gamma \left(\frac{1}{2} (2 d-5)\right)b^{6-2 d}}{2\Gamma (d-2)} \Bigg((d-1) \Big(2\log (b)+\psi ^{(0)}(d-2)\Big) \nonumber\\
&+&(1-d)\psi ^{(0)}\left(\frac{2d-5}{2}\right)-\frac{(\omega+2)}{\omega}\Bigg),\\
B & = & \frac{\Gamma \left(\frac{d-2}{2}\right)b^{3-d}}{\Gamma \left(\frac{d-1}{2}\right)}\Bigg( 2(d -2)\Big(\log (b)-\frac{1}{\omega(d-1)}-\frac{1}{(d -2)}\Big)\nonumber\\
&+&\Bigg(\psi ^{(0)}\left(\frac{d-1}{2}\right)- \psi ^{(0)}\left(\frac{d-2}{2}\right)\Bigg) \Bigg). \nonumber
\end{eqnarray}
Here,  $\psi$ is  the polygamma function defined as follows
\begin{equation}
\label{69}
\psi^{n}(z)={\frac  {d^{n}}{dz^{n}}}\psi (z)={\frac  {d^{{n+1}}}{dz^{{n+1}}}}\ln \Gamma (z), \quad
\psi^{(0)}(z)=\psi(z)=\frac{\Gamma'(z)}{\Gamma(z)}.
\end{equation}
Having computed   the    deflection angle, we move to analyse and discuss the associated behavior.  In particular, we consider the effect of   the  impact parameter $b$, DE, and the dimension on such quantity.    In  Fig.\ref{f4}, we  represent, indeed,  the variation of the  deflection  angle $\Theta$  as a function of the parameter $b$ in different dimensions $d$ with DE. It follows from such a  figure,  based on the top panels one,  that $\Theta$ decreases by increasing the impact parameter $b$ and it  increases by  increasing  the field intensity $c$.
In the two bottom panels, we observe that the deflection angle $\Theta$ decreases gradually when the dimension of the space-time $d$ increases.

\begin{figure}[ht!]
\begin{center}
\includegraphics[width=16cm, height=5.cm]{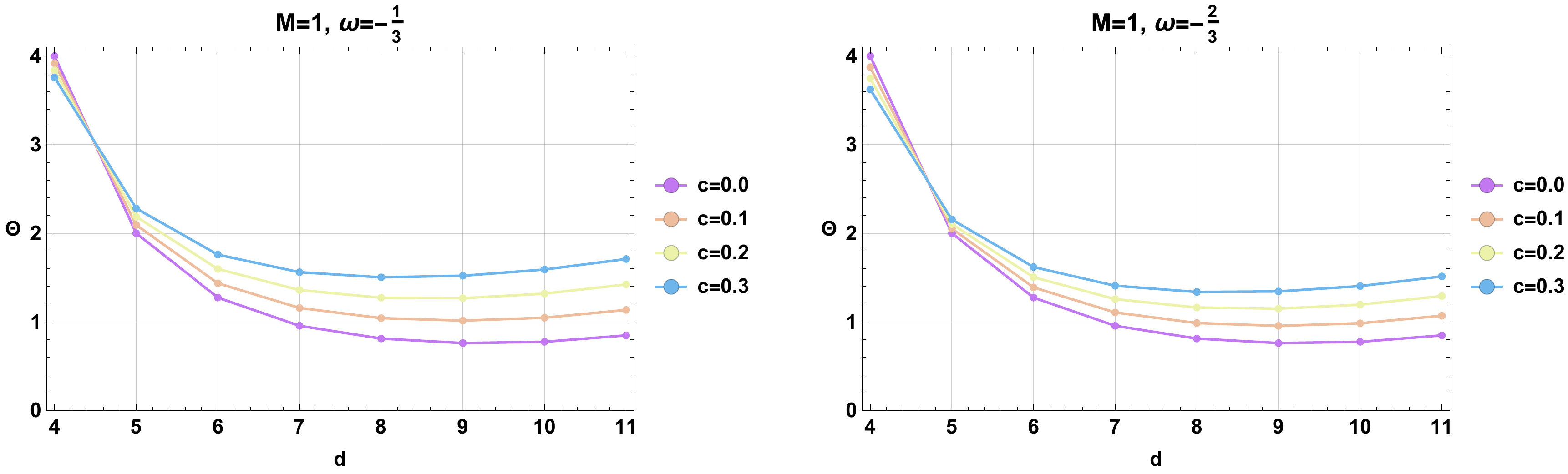}
\end{center}
\caption{\footnotesize \it Variation of $\Theta$ as function of  the space-time dimensions $d$ for fixed parameter $b=1$. }
\label{f5}
\end{figure}

To better  visualize such a behavior,  we plot in Fig.\ref{f5} the variation of $\Theta$ in terms of  the space-time dimensions $d$ for fixed values of the  impact parameter $b$.

It follows from this figure that the effect of DE  on $\Theta$ becomes relevant from $d=5$. Such an angle, being relevant   in $d=4$,  decreases with space-time dimension $d$. However,  for higher dimensions it is  almost constant. Its value increases with the  DE intensity  field $c$. This  visible  behavior seems to have possible connections  with  ideas  corresponding   to  DE as  the extra dimension evidence  predicted by  M-theory and superstring models.

\section{Optical behaviors from the charge effect}
\begin{figure}[!ht]
		\centering
		\hspace{-1.7cm}
		 \includegraphics[scale=.4]{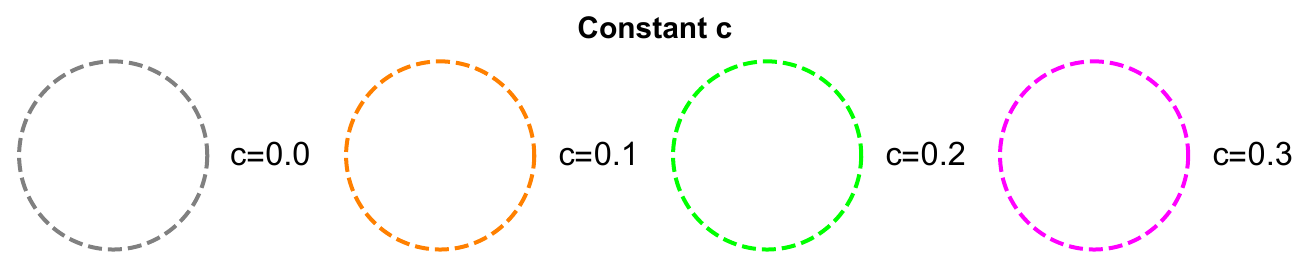}
			\begin{tabbing}
			\hspace{5.2cm}\= \hspace{5.2cm}\=\kill
			\includegraphics[scale=.39]{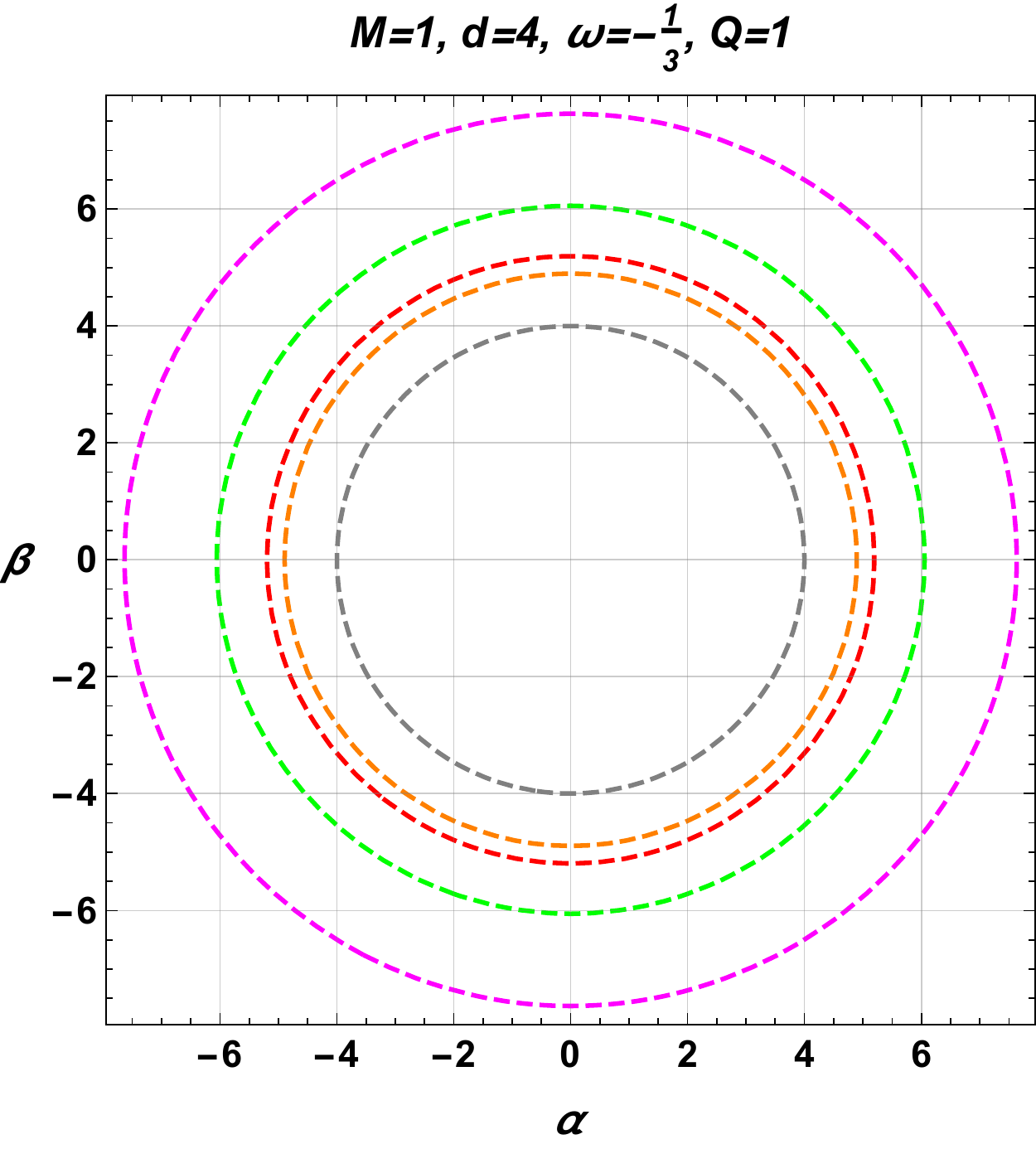} \>
			\includegraphics[scale=.39]{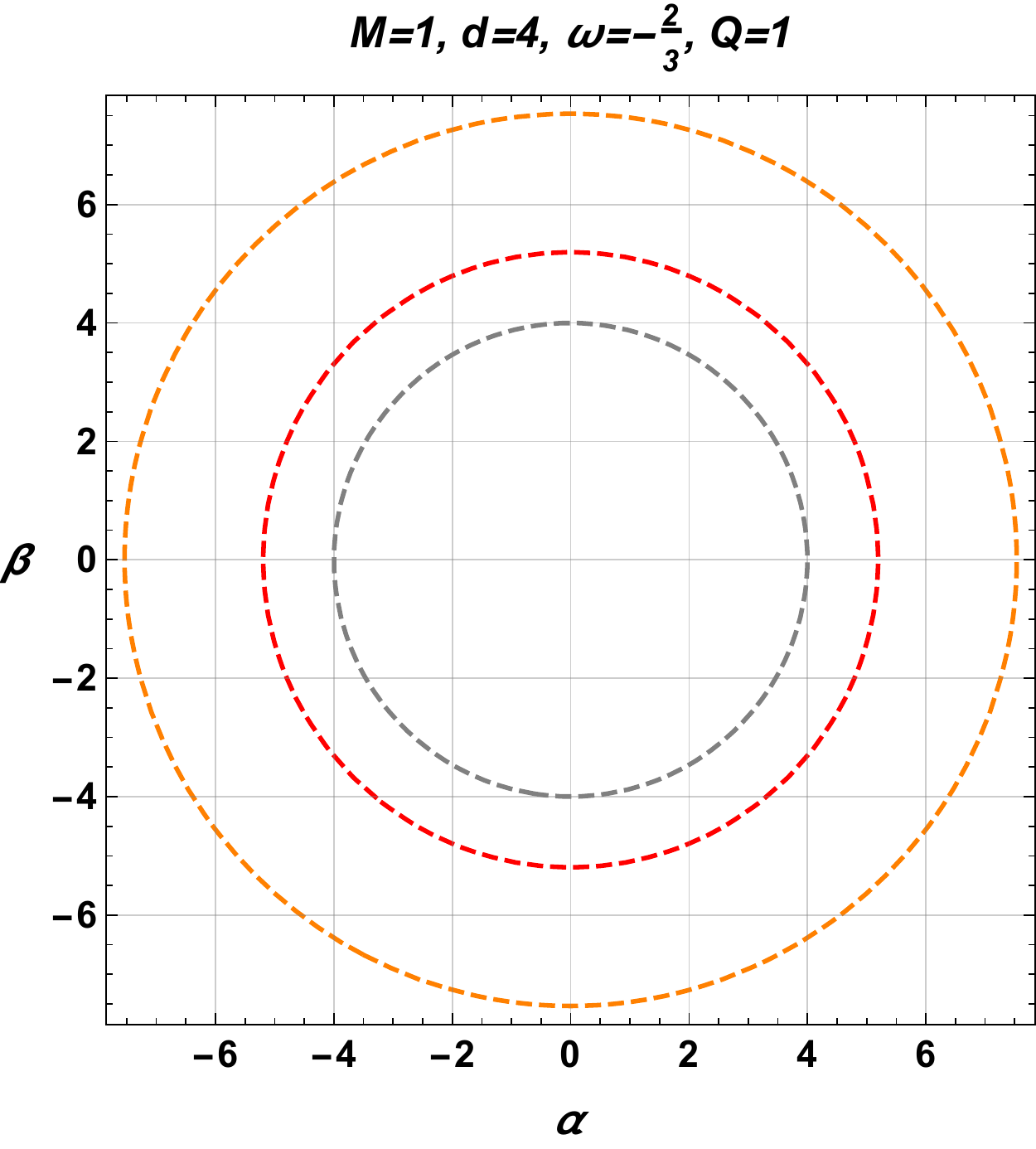} \>
			\includegraphics[scale=.39]{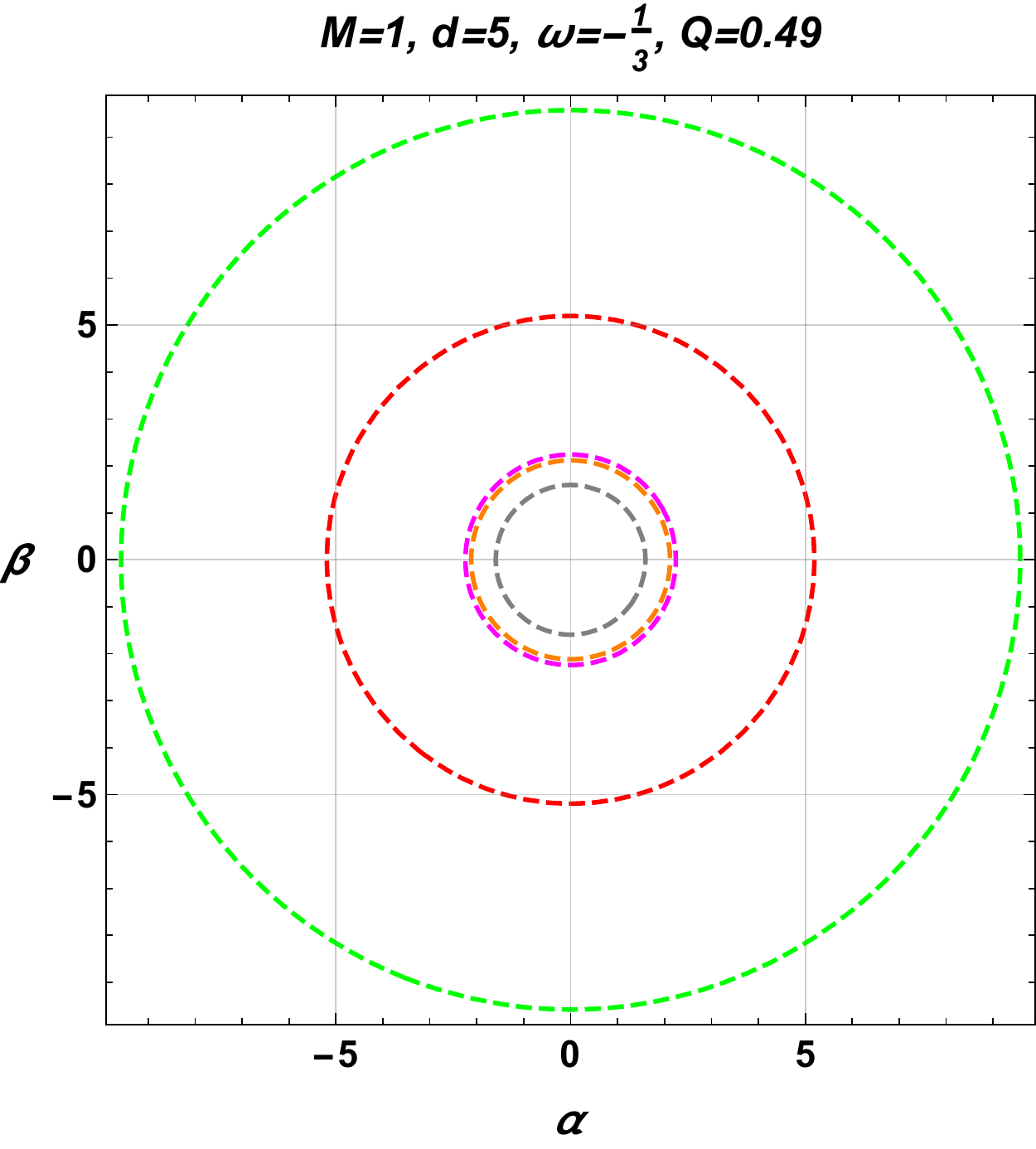} \\
			\includegraphics[scale=.39]{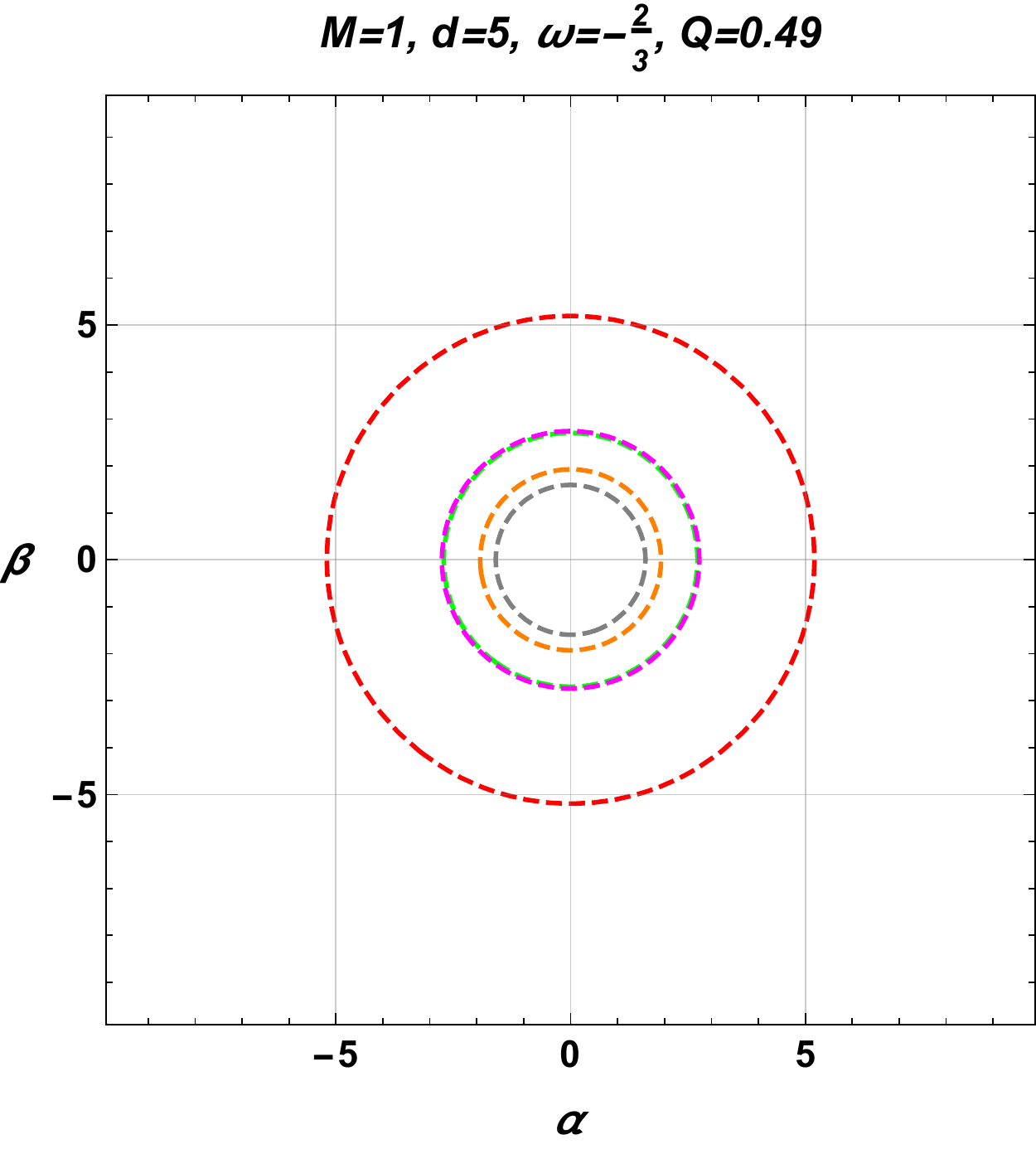} \>
			\includegraphics[scale=.39]{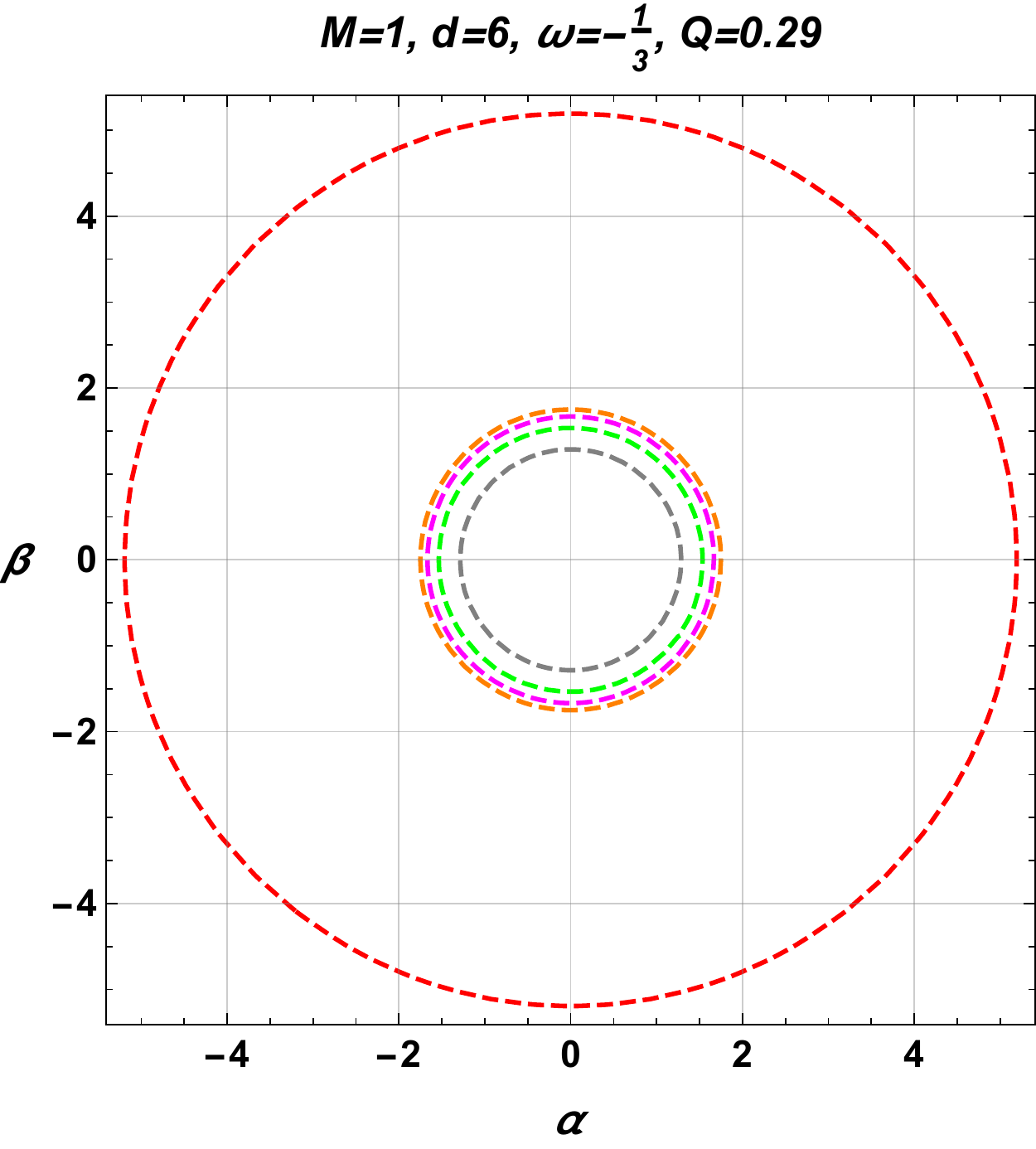} \>
			\includegraphics[scale=.39]{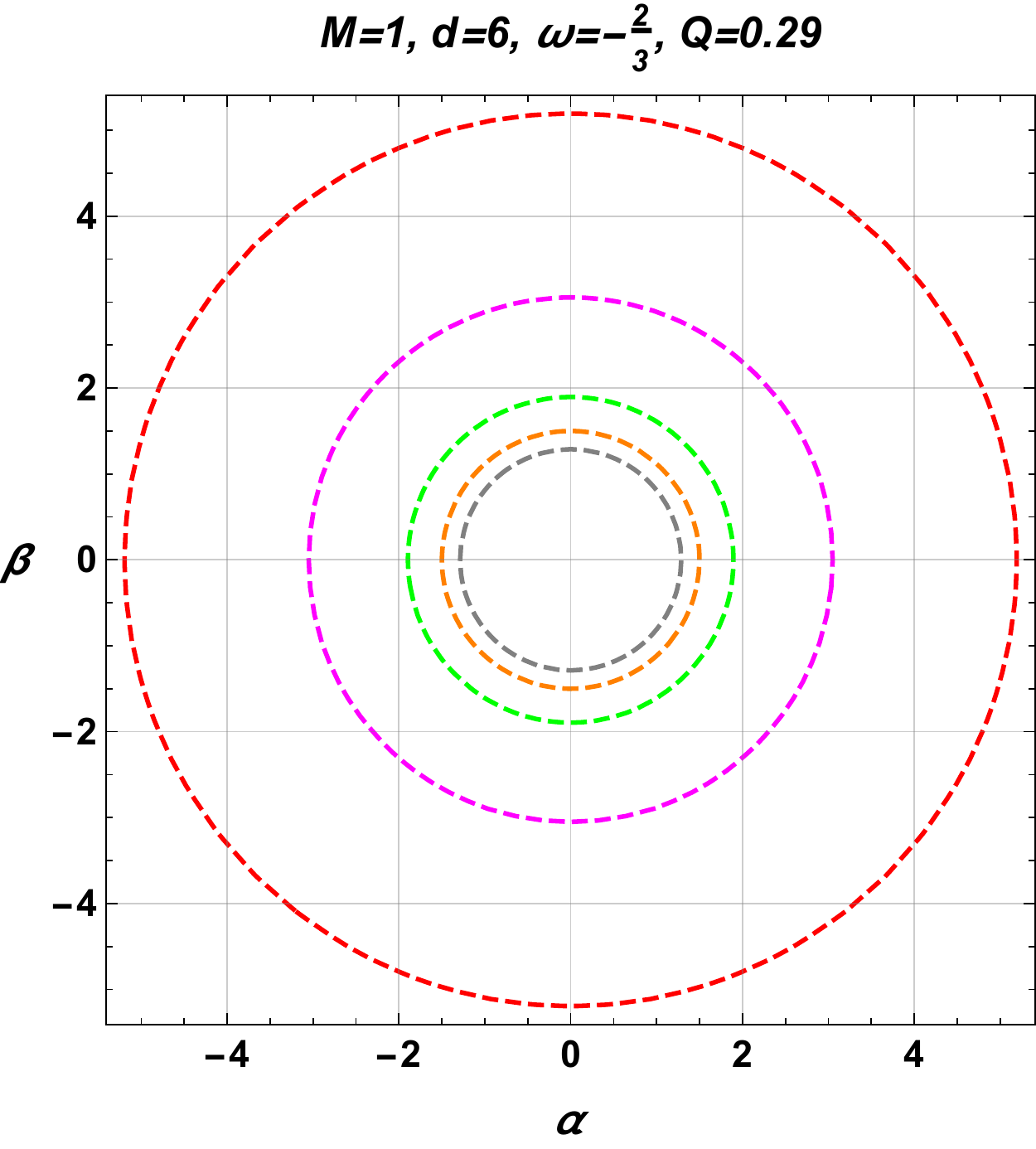} \\
			 \hspace{2.5cm}
			\includegraphics[scale=.39]{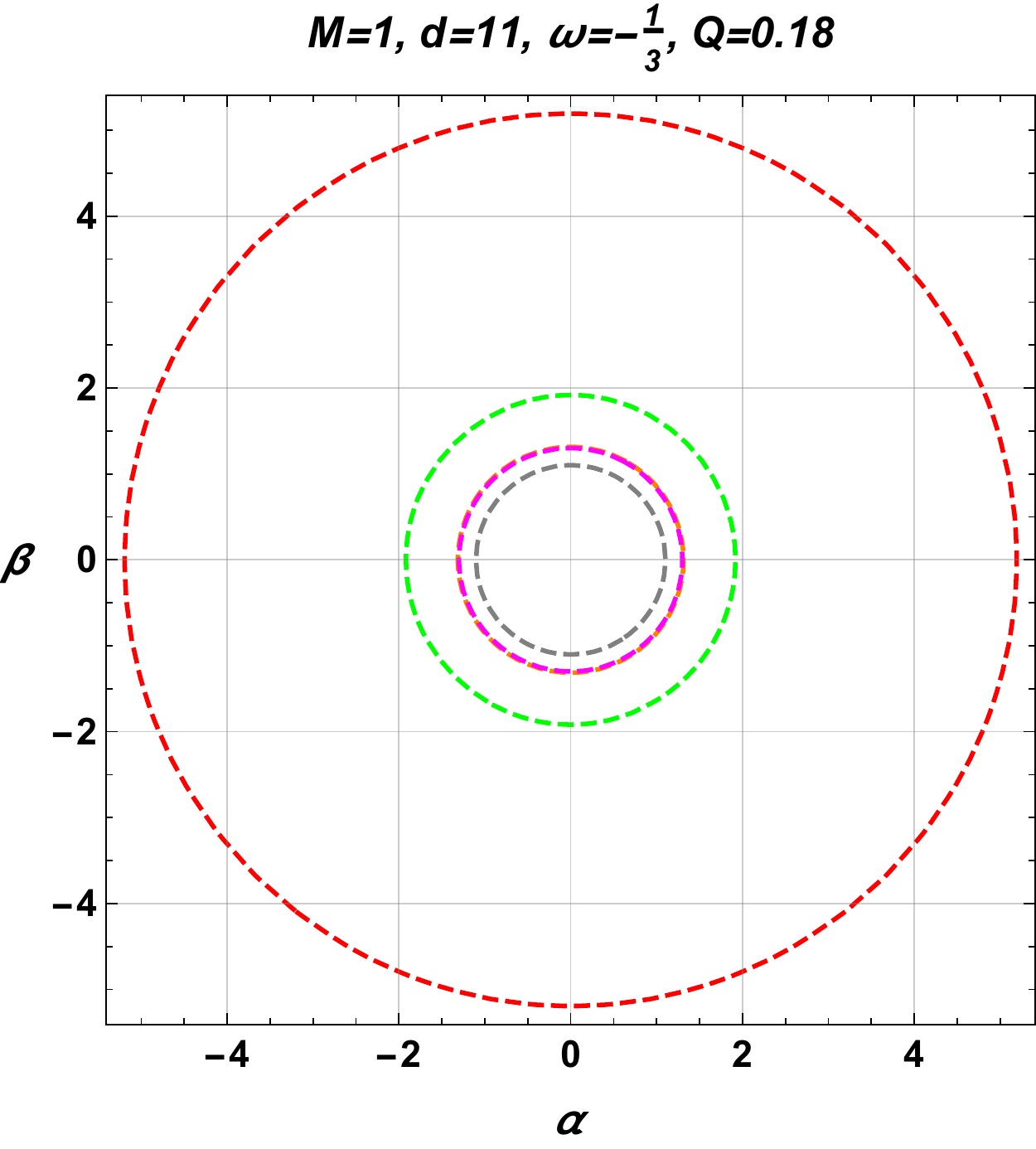} \>
			 \hspace{2.5cm}
			\includegraphics[scale=.39]{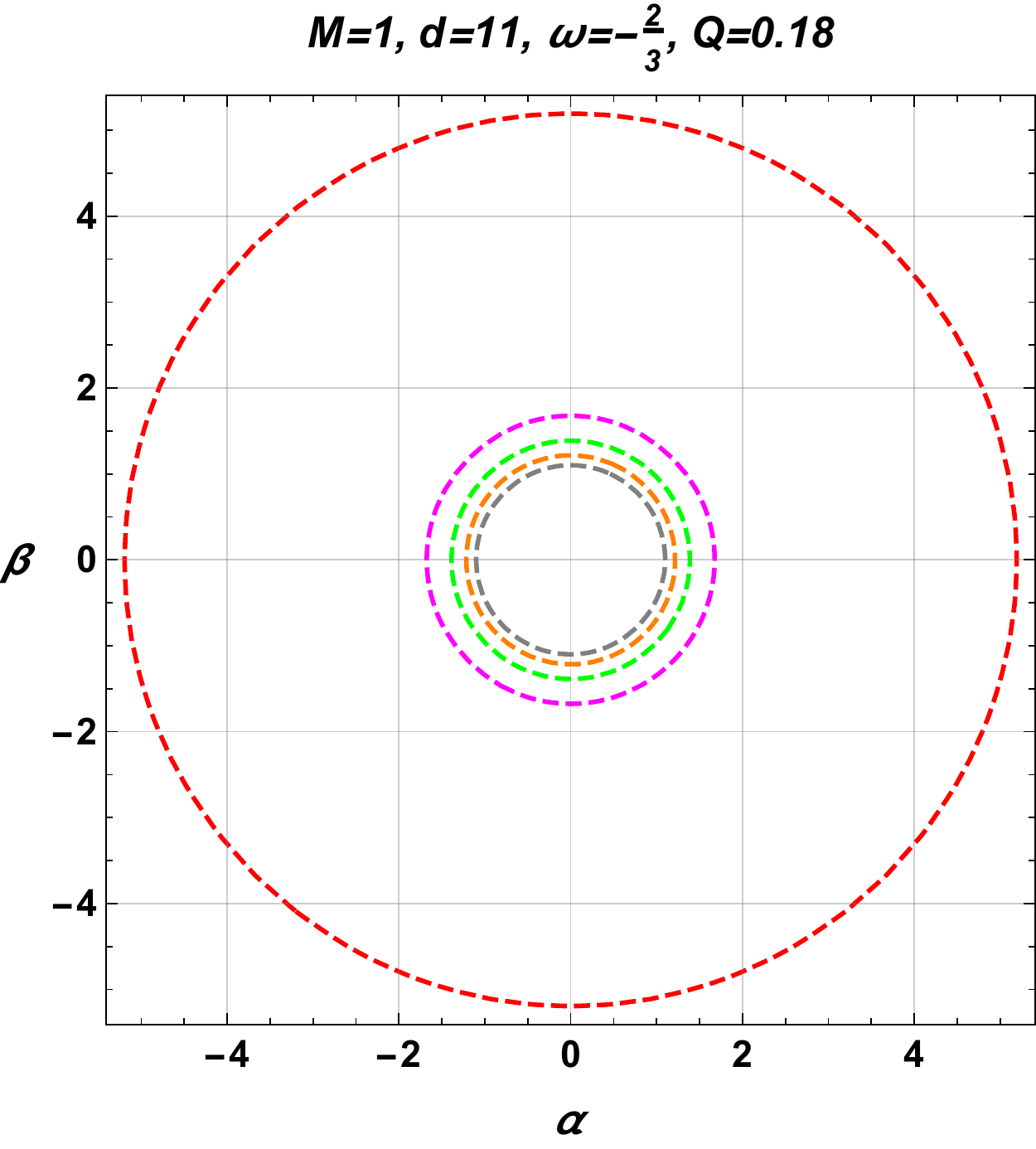} \\
				        \end{tabbing}

\caption{\footnotesize \it Shadow of charged black hole with DE $(-\frac{1}{3})$-model and $(-\frac{2}{3})$-model in different dimension $d$ as a function of $c$. In the all panels,  the red circle corresponds to the Schwarzschild black hole shadow without DE and charge $ (d=4, \ c=0, \ Q=0)$.}
\label{g33}
\end{figure}

 In this section, we unveil more  behaviors by  introducing the  charge effect. To start, the associated  metric function $f_\omega(r)$ reads as  \begin{equation}\label{chargefr}
f_{\omega}(r)=1-\frac{\mu}{r^{d-3}} + \frac{Q^2}{r^{2(d-3)}} - \frac{c}{r^{\omega(d-1)+d-3}},
\end{equation}
where $Q$ is the charge of the black hole.  It has been observed that  for  $d>4$,   the solution of the associated  equation  generates  two radius photon spheres $r_0$(max) and $r_0$(min). This gives   two radius of the  shadow circles. However, to deal with  the shadow behaviors, one should  use the maximum one, as shown   in  Fig.\ref{g33}. It has been remarked that the charge increases the shadow size for fixed   values of $c$ and $\omega$.  A close inspection shows that non  trivial behaviors arise for the  charged case compared to  the results of the non-charged black hole shadows presented in Fig.\ref{f2}. In  $d=4$,  we observe that  for constant values of the  charge and for $d>4$, $c=0$ or $c>0.2$, the shadow size decreases.   However, for   $d>4$ and $0<c<0.2$ such behaviors depend strongly on the charge where  the shadow size increases. This is probably due to the competition  between the positive charge therm and the negative DE term in the blacking function in Eq.\eqref{chargefr}.  We expect also that such behaviors  could be related to  extra dimensions  supporting the discussion  of DE from many aspects.

To go deeply in  such an  analysis related to  the charge black hole effect, we use the same  precedent procedure to evaluate the deflection angle behaviors.
Using  \eqref{42} and   the metric  function $f_{\omega}(r)$ of  the charged black solution, we get the optical metric  in the  higher dimensional space-time. Similar calculations show that  the Gaussian curvature of the optical charged black hole up to leading orders ($\mathcal{O}$($M^2$,$c^2$))  can be given by
\begin{equation}
\label{d1}
\begin{split}
\mathcal{K}\approx&\left(2 d^2-11 d+15\right) Q^2 r^{4-2 d}-\frac{ \mu \left(d^2-5 d+6\right)}{2}\left(3 Q^2 r^{7-3d}+r^{1- d}\right)\\
&-\frac{1}{2r^{\omega (d-1)}}\Bigg( \Big(\left(d^2-2 d+1\right) \omega ^2+\left(2 d^2-7 d+5\right) \omega +\left(d^2-5 d+6\right)\Big)r^{1-d}\\
&+Q^2\Big(\left(d^2-2 d+1\right) \omega ^2+(d-1)\omega+\left(3 d^2-15 d+18\right) \Big)r^{7-3d}\\
&-\mu\left(d^2-2 d+1\right) \omega ^2+\left(d^2-3 d+2\right) \omega +\left(d^2-4 d+3\right)r^{4-2d}\Bigg)c.
\end{split}
\end{equation}
It is  remarked   that  this result recovers the  Gaussian curvature of optical four  dimensional charged black hole presented  in \cite{m2} and the higher dimensional case  investigated  in the previous section.  Exploiting  a power series expansion, we  obtain   the deflection angle
\begin{equation}
\label{ }
\begin{split}
\delta\approx&\Theta-\frac{Q^2}{12} \sqrt{\pi } b^{6-3 d}\Bigg(\frac{12 b^d \Gamma \left(\frac{2d-3}{2}\right)}{\Gamma (d-2)}+\frac{b^3}{\Gamma \left(\frac{3 d-7}{2}\right)}\Bigg(c\Big(D+E+F\Big)+6 (2-d) \mu\Bigg)\Bigg)\\
&+\mathcal{O}(M^2,c^2,\omega^2),
\end{split}
\end{equation}
where one has
\begin{eqnarray}
D & = & d^2 \left(6 \epsilon  \log (b)-3 \epsilon  \psi ^{(0)}\left(\frac{3 d}{2}-4\right)+3 \epsilon  \psi ^{(0)}\left(\frac{1}{2} (3 d-7)\right)\right), \\
E & = & d \left(-18 \epsilon  \log (b)+9 \epsilon  \psi ^{(0)}\left(\frac{3 d}{2}-4\right)-9 \epsilon  \psi ^{(0)}\left(\frac{1}{2} (3 d-7)\right)+2 \epsilon -6\right),\\
F & = &  \left( 12 \epsilon  \log (b)-6 \epsilon  \psi ^{(0)}\left(\frac{3 d}{2}-4\right)+6 \epsilon  \psi ^{(0)}\left(\frac{1}{2} (3 d-7)\right)-2 \epsilon +12\right),
\end{eqnarray}
 and  where $\Theta$  is  nothing but the deflection angle of the  non-charged QBH \eqref{68}. For $d=4$, we recover  the same result for the charged black hole \cite{m20,m2}.

\begin{figure}[!ht]
		\begin{center}
		\centering
			\begin{tabbing}
			\centering
			\hspace{9.3cm}\=\kill
			\includegraphics[scale=.5]{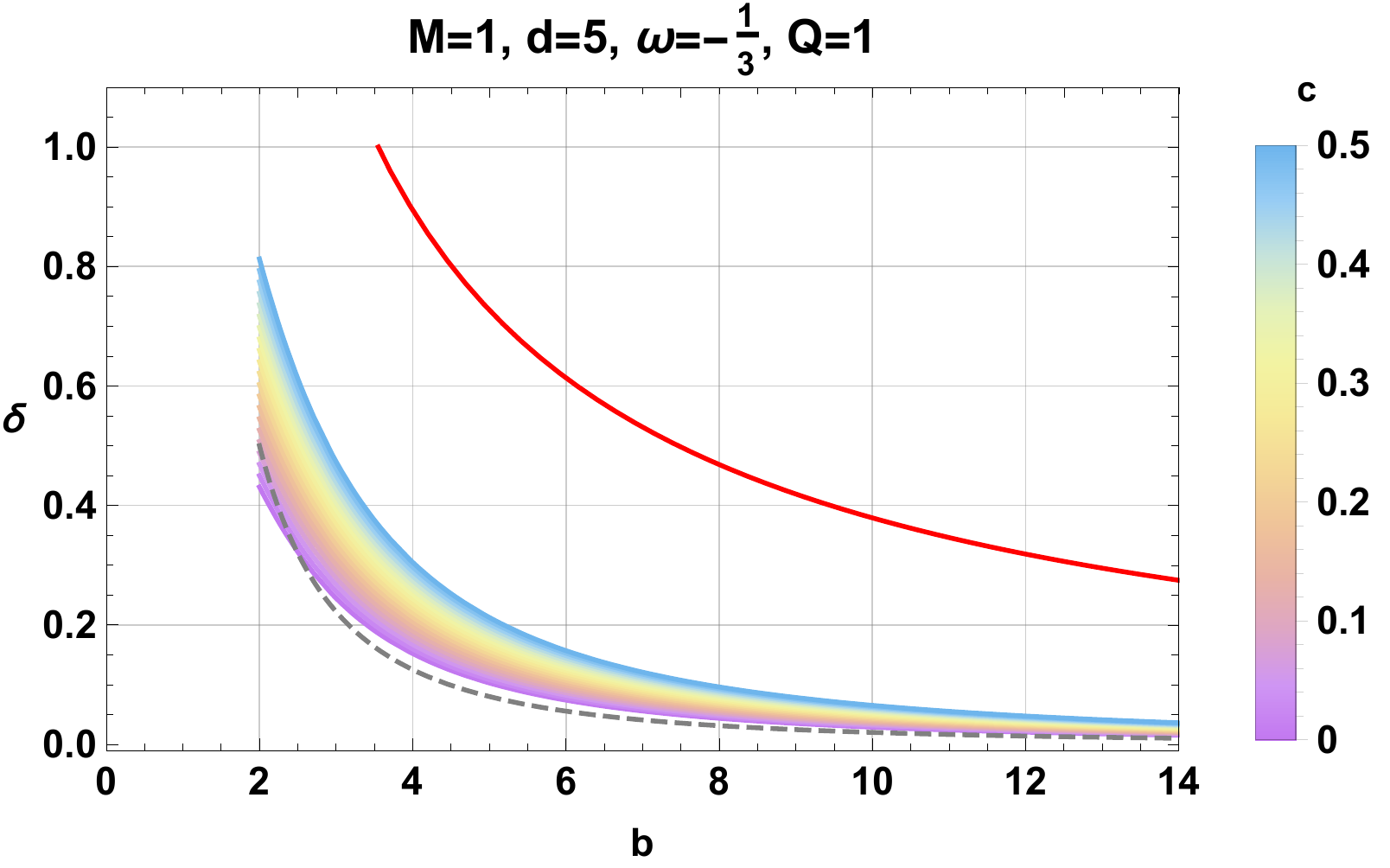} \>
			\includegraphics[scale=.5]{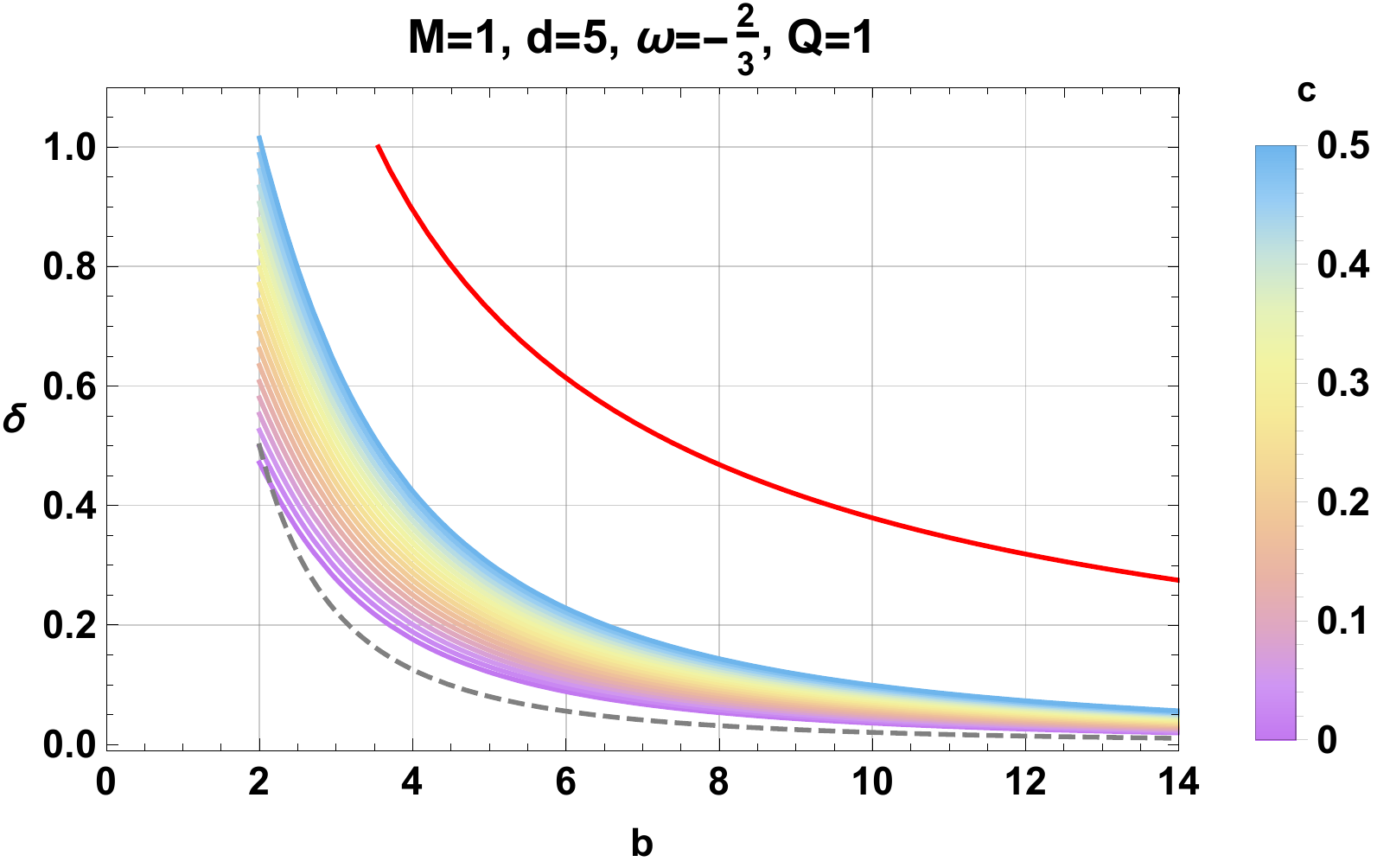} \\
			\includegraphics[scale=.5]{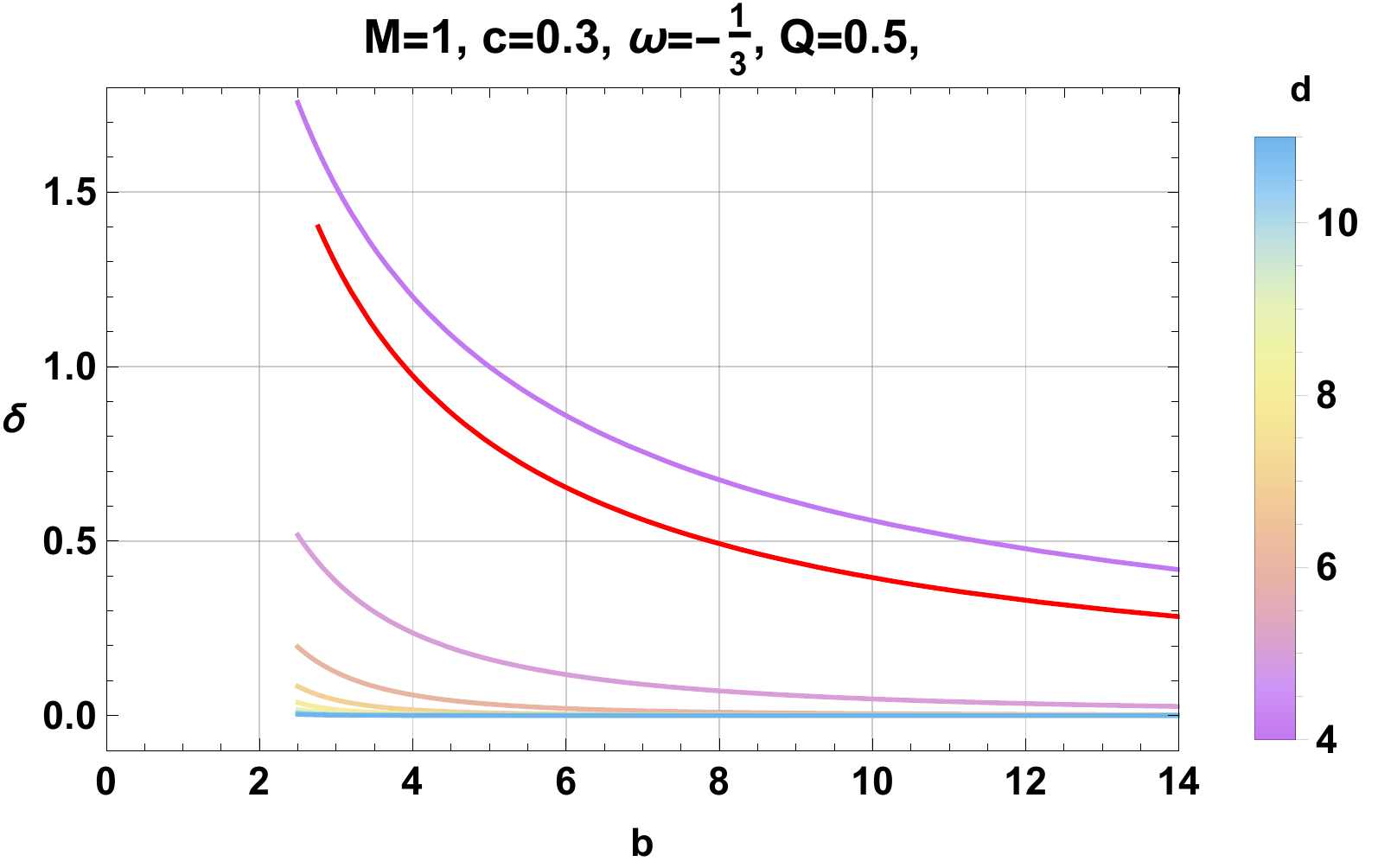} \>
			\includegraphics[scale=.5]{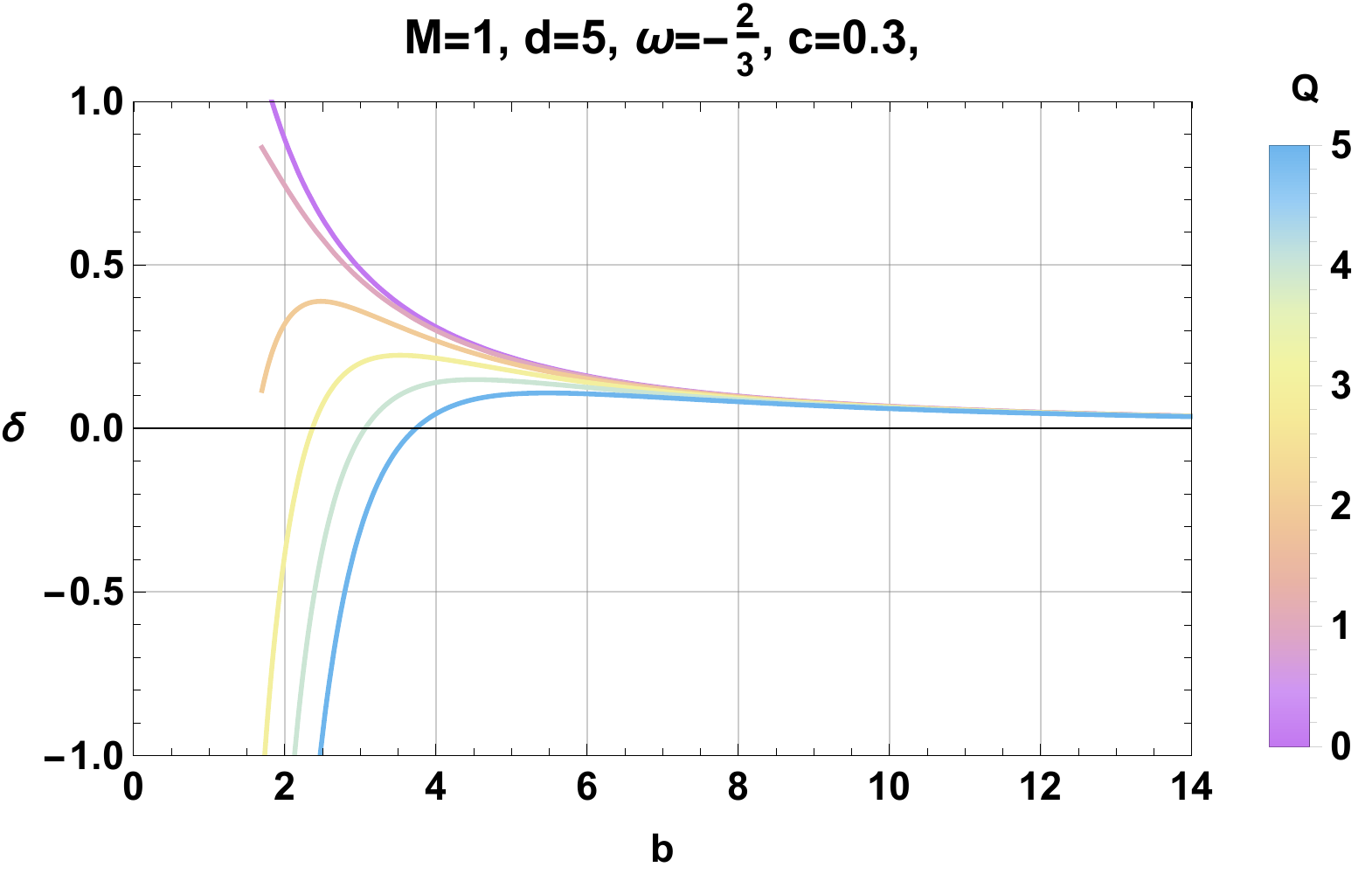} \\
	        \end{tabbing}

\caption{\footnotesize \it Variation of $\delta$ as a function of dimensions $d$ for different values of $c$ and  for fixed parameter $b$ and $Q$. In the all panels,  the gray  dashed line curve corresponds to the Schwarzschild black hole shadow without DE and charge in $d=5$.}
\label{g3}
\end{center}
\end{figure}

 In  Fig.\ref{g3}, we illustrate the charge effect on  the deflection angle.
For a fixed charge value,   the variation of  $\delta$ in terms of the impact parameter and the field intensity $c$  remain the same as the neutral case. Thus means that   $\delta$ decreases gradually within $b$. However,  we can easily notice that  the growth of $c$ increases $\delta$. The left right panel  indicates that when the charge $Q$ increases the deflection angle $\delta$ decreases.
\section{Shadow behavior in the presence of the plasma }
The present  approach can be  adaptable to a broad variety of   backgrounds. It   has been  remarked  that  the  previous  space-time can be modified  by  a non-trivial background   associated with a  non-magnetized cold plasma \cite{m23}. According to \cite{m21,m22,m24},  the frequency $\omega_p$ of the electron plasma can be written,  using   a radial power law, as follows
\begin{equation}
\label{p1}
 \omega_{p}^2(r)=\frac{k}{r^h},  \qquad  h \geq 0.
\end{equation}
 Using  the plasma frequency  $\omega_p$ and the  photon frequency $\omega_0$,  the refraction index  of such  a background reads as
\begin{equation}
\label{p3}
n^2(r,\omega_0)= 1- \frac{\omega^2_p(r)}{\omega^2_0}.
\end{equation}
In this situation,  certain  vacuum equations  including  the   equations of motion for  photon around  QBH  can be
 modified. In particular, the relevant ones   become
\begin{eqnarray}
\frac{dt}{d\tau}& = & \frac{n^2(r,\omega_0)E}{f_{\omega}(r)},\\
\mathcal{R}(r)& = &n^2(r,\omega_0)E^2r^4-r^2f_{\omega}(r)(\mathcal{K}+L^2).
\end{eqnarray}
Using   Eq.\eqref{24},  the impact parameters $\eta$ and $\xi$   can be generalized to
\begin{equation}
\label{p9}
\eta+\xi^2=\frac{5n^2(r_0,\omega_0)r_0^2+2n(r_0,\omega_0)n'(r_0,\omega_0)r^3_0}{3f_\omega(r_0)+rf_\omega'(r_0)},
\end{equation}
where  the prime indicates  the derivative with respect to $r$. Redefining  the  celestial coordinates  \eqref{35} associated with  the  equatorial hypeplan   as follows
\begin{equation}
\label{p10}
\alpha=-\frac{\xi}{n(r_0,\omega_0)}, \hspace{1cm} \beta=\pm\frac{\sqrt\eta}{n(r_0,\omega_0)},
\end{equation}
the equation  \eqref{p9}  can be  reexpressed as
\begin{equation}
\label{p11}
\alpha^2+\beta^2=\frac{\xi^2+\eta}{n^2(r_0,\omega_0)}=\frac{5n^2(r_0,\omega_0)r_0^2+2n(r_0,\omega_0)n'(r_0,\omega_0)r^3_0}{n^2(r_0,\omega_0)\big(3f_\omega(r_0)+rf_\omega'(r_0)\big)}.
\end{equation}

\begin{figure}[!h]
\begin{center}
\includegraphics[width=16cm, height=4.cm]{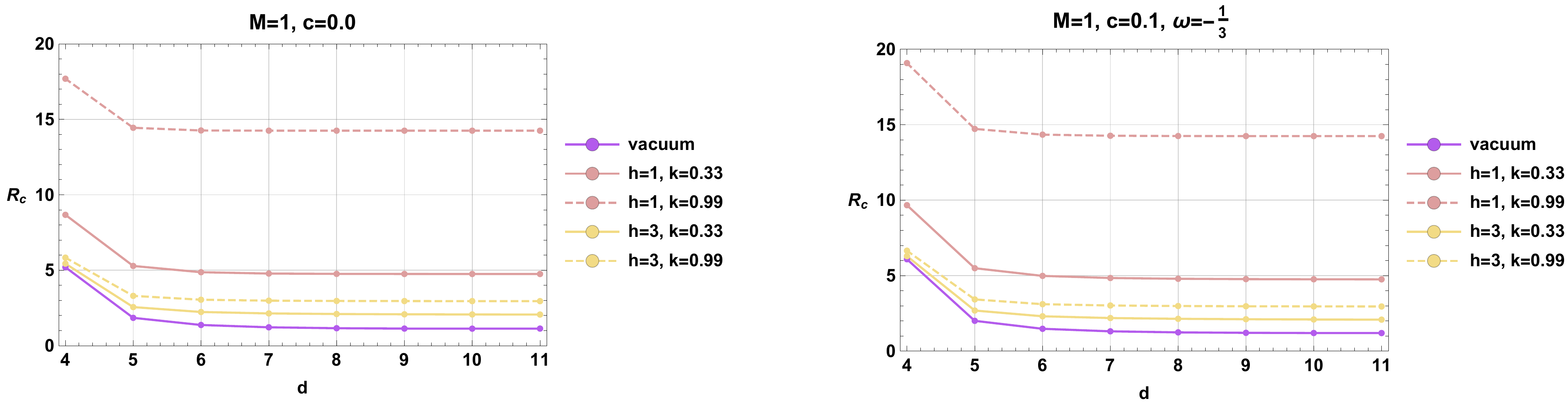}
\end{center}
\caption{\it \footnotesize Variation of shadow radius $R_c$  as a function dimension $d$,  without DE and for the $(-\frac{1}{3})$-model for  different values of the intensity $h$ and $k$.}
\label{fp5}
\end{figure}
 Fig.\ref{fp5} shows  the variation of the shadow radius in the presence of plasma as a function of the space-time  dimension $d$ with and without DE.
Turning off  DE, we have  observed that the plasma presence increases the shadow radius $R_c$.  For $h=1$,   we have remarked  also that $R_c$ increases when $k$ increases. The implementation of  DE, however,  has  revealed that the shadow radius increases only for $ 4 \leq d \leq 6$. The higher dimensional cases seem to have approximatively the ordinary behaviors.  We expect that other optical properties could be also approached  in such plasma backgrounds by performing  similar calculations.
\pagebreak
\pagebreak
\newline
\section{Conclusions and open questions}
Motivated by M-theory/superstring inspired models, we have investigated certain behaviors of  the deflection angle and shadow  circular  shapes of higher dimensional QBH associated with $\omega=-\frac{1}{3} $ and  $\omega=-\frac{2}{3} $ models.  In particular, we have  derived the complete geodesic structure of photons around such  black holes using the Hamilton-Jacobi equation and Carter's constant separable method. Linking   the celestial coordinate  to the   geodesic equations and plotting  the black hole shadow shape within the field intensity $c$ and the dimension of the space-time $d$, we have found that  $d$ decreases the shadow size. However,   DE increases  such  a geometrical size.  Moreover, we have  computed  the energy emission rate of the black hole by assuming that the area of the photon sphere is equal to the high-energy absorption cross-section.  Then,  we  have analyzed  the effect of the DE field  intensity   $c$ and the space-time dimension $d$ on such a quantity.
In the second part of the present work, we  have approached  the weak deflection angle of light of such  higher dimensional QBH.  Precisely, this has been done by  determining the corresponding  optical Gaussian curvature. Using the Gauss-Bonnet theorem associated with   the optical metric,  we have calculated  the leading terms of the deflection angle in the weak-limit approximation.  Besides, we have discussed  the impact  of  DE and the space-time dimension $d$  on  such a optical quantity. In the last part, we have  analyzed  the effect of the  charge $Q$ on all the above-computed  quantities,  which not only   provides   non-trivial behaviors but also recovers the ordinary ones.

In a  arbitrary dimensional space-time,  it  has been  shown that the quintessence field increases the size of shadows. According  to four-dimensional results reported in \cite{TaoTao1}, we could expect  that such a field imposes an impact on the size of the image of black hole and the distances between the observer and the horizon of the event  through the   state parameter. In this way,  the positions of the photon spheres of the image  could be modified. In future EHT experiments,   the observations of black hole shadow images could  provide insights and  certain signatures associated with  the quintessential dark energy physical  models. \\
In   coming  works,  we attempt   to study the impact of the equivalency principle on the black hole in the presence of quintessence fields by incorporating  the spinning parameter  explored in \cite{TaoTao2}. We anticipate that the principle of equivalence could remain in the case of the non-rotating black hole in the presence of quintessence fields.  However, the rotating parameter may bring non-trivial results, which will be explored elsewhere.

This work comes up with many questions. The natural one is  to make contact with evidence of DE from extra dimensions. We hope to  address such a question in future  by considering M-theory analysis associated with the axionic  universe. Connections with Dark Matter could be also possible \cite{TaoTao}.  Moreover,  based on  the event telescope and  black hole investigations,  one could say   that one involves  new and powerful tools to  approach  the so  called new  physics beyond standard model.

%
%

\section*{Acknowledgment}
  This work is partially supported by the ICTP through AF-13. We are grateful to the anonymous referees for their careful reading of our manuscript, insightful comments, and suggestions, which have allowed us to improve this paper significantly.
\appendix
\section{Einstein equations in $d$-dimensional  space-time}
In this appendix, we  collect    some calculations associated with the higher dimensional  quantities used  trough this work.
\subsection*{Christoffel symbols}
\begin{equation}
\begin{split}
     & \Gamma^t_{t\,\,r} \;=\;\frac{ \nu'}{2},  \\
      &\Gamma^r_{t\,\,t}\;=\;\frac{ \nu' e^{\nu-\lambda}}{2},\\
        &\Gamma^{r}_{\theta_i\,\,\theta_i}\;=\;-re^{-\lambda}\prod_{j=1}^{i-1}\sin^2{\theta_j}, \hspace{1cm} i=1,\ldots, d-2,\\
        & \Gamma^{\theta_i}_{r\,\,\theta_i}\;=\;\frac{1}{r}, \hspace{1cm} i=1,\ldots, d-2,\\
       &\Gamma^{\theta_k}_{\theta_i\,\,\theta_i}\;=\;-\cos{\theta_k}\sin{\theta_k}\prod_{j=k+1}^{i-1}\sin^2{\theta_j}, \hspace{1cm} k=1,\ldots, d-3, \hspace{0.5cm} i=k+1, \ldots, d-2,\\
         &\Gamma^{\theta_k}_{\theta_i\,\,\theta_k}\;=\;\cot\theta_i, \hspace{1cm} k=2,\ldots, d-2, \hspace{0.5cm} i=1,\ldots, k-1.
\end{split}
\end{equation}
where $\nu'=\frac{\partial \nu}{\partial r}$ and $\lambda'=\frac{\partial \lambda}{\partial r}$.
\subsection*{Riemann tensor}
\begin{equation}
\begin{split}
&\mathrm{R}^t_{r\,\,r\,t}\;=\;\frac{-\lambda'\nu'+\nu'^2+2\nu''}{4},\\
&\mathrm{R}^t_{\theta_i\,\,\theta_i\,t}\;=\;\frac{r\nu'e^{-\lambda }}{2}\prod_{j=1}^{i-1}\sin^2\theta_j,  \hspace{1cm}  i=1,\ldots, d-2,\\
&\mathrm{R}^r_{\theta_i\,\,\theta_i\,r}\;=\;-\frac{r\lambda'e^{-\lambda }}{2}\prod_{j=1}^{i-1}\sin^2\theta_j,  \hspace{1cm}  i=1,\ldots, d-2,\\
&\mathrm{R}^r_{t\,\,r\,t}\;=\;\frac{e^{\nu-\lambda}(-\lambda'\nu'+\nu'^2+2\nu'')}{4},\\
&\mathrm{R}^{\theta_i}_{t\,\theta_i\,t}\;=\;\frac{e^{\nu-\lambda}\nu'}{2r},  \hspace{1cm}  i=1, \ldots, d-2,\\
&\mathrm{R}^{\theta_i}_{r\,\theta_i\,r}\;=\;\frac{\lambda'}{2r},  \hspace{1cm}  i=1,\ldots (d-2),\\
&\mathrm{R}^{\theta_i}_{\theta_k\,\theta_k\,\theta_i}\;=\;(e^{-\lambda}-1)\prod_{j=1}^{k-1}\sin^2\theta_j, \hspace{1cm} i=1,\ldots, d-3, \hspace{0.5cm} k=i+1, \dots, d-2,\\
&\mathrm{R}^{\theta_i}_{\theta_k\,\theta_k\,\theta_i}\;=\;(1-e^{-\lambda})\prod_{j=1}^{k-1}\sin^2\theta_j, \hspace{1cm} i=2,\ldots, d-2, \hspace{0.5cm} k=1,\ldots, i-1.
\end{split}
\end{equation}
\subsection*{ Ricci tensor}
\begin{equation}
\begin{split}
&{R_{t\,\,t}}\;=\;\frac{e^{\nu-\lambda}(\nu'((2d-4)-r\lambda')+r\nu'^2+2r\nu'')}{4r},\\
&{R_{r\,\,r}}\;=\;\frac{\lambda'((2d-4)+r\nu')-r(\nu'^2+2\nu'')}{4r},\\
&{R_{\theta_i\,\theta_i\,}}\;=\;\frac{e^{-\lambda}}{2}((2d-6)(e^{-\lambda}-1)+r(\lambda'-\nu'))\prod_{j=1}^{i-1}\sin^2\theta_j, \hspace{1cm} i=1,\ldots, d-2.
\end{split}
\end{equation}
\subsection*{Scalar curvature}
\begin{equation}
R\;=\;\sum^{d-2}_{i=1}\frac{(r(\nu'-\lambda')+(2d-6)(1-e^\lambda))e^{\lambda}}{2r^2} \,+\, \frac{e^{-\lambda } \left((d-2) \nu '-\lambda ' \left(d+r \nu '-2\right)+2 r \nu ''+r \left(\nu '\right)^2\right)}{2 r}.
\end{equation}
\subsection*{Einstein tensor}
\begin{equation}
\begin{split}
&{G_{t\,\,t}}\;=\;\frac{(d-2) e^{\nu -\lambda } \left((d-3) \left(e^{\lambda }-1\right)+r \lambda '\right)}{2 r^2},\\
&G_{r\,\,r}\;=\;\frac{(d-2) \left((d-3) \left(1-e^{\lambda }\right)+r \nu '\right)}{2 r^2},\\
&{G_{\theta_i\,\theta_i\,}}\;=\;\frac{1}{4} e^{-\lambda } \left(2 (d-3) r \left(\nu '-\lambda '\right)+2 (d-4) (d-3)-r^2 \lambda ' \nu '+2 r^2 \nu ''+r^2 \left(\nu '\right)^2\right)\\
&\hspace{1.55cm}-\frac{1}{2} (d-3) (d-4) \prod_{j=1}^{i-1}\sin^2\theta_j, \hspace{1cm} i=1,\ldots, d-2.
\end{split}
\end{equation}
\subsection*{Einstein equation}
In the reduced units,  we give
\begin{equation}
\begin{split}
&2T_{t}^{\;t}\;=\frac{(d-2) e^{-\lambda } \left((d-3) \left(e^{\lambda }-1\right)+r \lambda '\right)}{2 r^2},\\
&2T_{r}^{\;r}\;=\;\frac{(d-2) e^{-\lambda } \left((d-3) \left(e^{\lambda }-1\right)-r \nu '\right)}{2 r^2},\\
&2T_{\theta_i}^{\;\theta_i}\;=\;-\frac{e^{-\lambda } \left(2 (d-3) r \left(\nu '-\lambda '\right)+2 (d-4) (d-3)-r^2 \lambda ' \nu '+2 r^2 \nu ''+r^2 \left(\nu '\right)^2\right)}{4 r^2}\\
&\hspace{1.55cm} + \frac{(d-3) (d-4)}{2 r^2}\hspace{1cm} i=1,\ldots, d-2.
\end{split}
\end{equation}

 \end{document}